\RequirePackage{fix-cm}
\documentclass[twocolumn,epjc3]{svjour3}  
\smartqed
\usepackage{amsmath}
\usepackage{amssymb}
\usepackage{nccmath}
\usepackage{graphicx}

\usepackage{ragged2e}
\usepackage{subcaption}
\usepackage{enumitem}
\usepackage{graphicx}
\usepackage{float}
\usepackage{hyperref}
\usepackage{multicol}
\usepackage{multirow}
\usepackage[switch]{lineno}  

\journalname{Eur. Phys. J. C}

\begin{document}
\title{Measurement of the absolute efficiency of the X--ARAPUCA photon detector for the DUNE Far Detector 1}

\author{
        R.~\'{A}lvarez-Garrote\thanksref{addr1}
        \and
        C.~Brizzolari\thanksref{addr2,addr3}
        \and 
        A.~Canto\thanksref{addr1}
        \and 
        E.~Calvo\thanksref{addr1}
        \and 
        C.M.~Cattadori\thanksref{addr2}
        \and 
        C.~Cuesta\thanksref{addr1}
        \and
        A.~de~la~Torre Rojo\thanksref{addr1}
        \and
        I.~Gil-Botella\thanksref{addr1}
        \and 
        C.~Gotti\thanksref{addr2}
        \and
        D.~Guffanti\thanksref{addr2,addr3}
        \and
        A.A.~Machado\thanksref{addr4.0}
        \and
        S.~Manthey Corchado\thanksref{addr1}
        \and 
        I.~Mart\'{i}n\thanksref{addr1}
        \and 
        C.~Massari\thanksref{addr2,addr3}
        \and
        L.~Meazza\thanksref{e2,addr2,addr3}
        \and 
        C.~Palomares\thanksref{addr1}
        \and 
        L.~P\'{e}rez-Molina\thanksref{e1,addr1}
        \and 
        E.~Segreto\thanksref{addr4}
        \and
        F.~Terranova\thanksref{addr2,addr3}
        \and
        A.~Verdugo~de~Osa\thanksref{addr1}
        \and
        H.~Vieira~de~Souza\thanksref{addr6} 
        \and
        D.~Warner\thanksref{addr5}
        }

\thankstext{e1}{e-mail: laura.perez@ciemat.es}
\thankstext{e2}{e-mail: luca.meazza@mib.infn.it}

\institute{
\textbf{CIEMAT}, Avda. Complutense~40, 28040 Madrid~(Spain) \label{addr1} 
\and
\textbf{Dipartimento di Fisica “Giuseppe Occhialini”, Università degli Studi di Milano-Bicocca}, Piazza della Scienza~3, 20126 Milano~(Italy) \label{addr2}
\and
\textbf{INFN Sezione di Milano-Bicocca}, Piazza della Scienza~3, 20126 Milano~(Italy) \label{addr3}
\and
\textbf{Dipartimento di Fisica, Università degli Studi Federico II}, Napoli 80126~(Italy) \label{addr4} 
\and
\textbf{Instituto de Física “Gleb Wataghin”, UNICAMP}, Campinas-SP, 13083-859~(Brazil) \label{addr4.0} 
\and
\textbf{CSU}, Limelight Ave, Castle Rock 80109~(United States) \label{addr5} 
\and 
\textbf{Laboratoire Astroparticule et Cosmologie}, rue Alice Domon et Léonie Duquet 10,  75013 Paris~(France)\label{addr6}
}

\date{Received: date / Accepted: date}

\maketitle

\begin{abstract}
The DUNE far detector has been designed to detect photons and electrons generated by the charged products of the interaction of neutrinos with a massive liquid argon (LAr) target. The Photon Detection System~(PDS) of the first DUNE far detector~(FD1) is composed of 6000 photon detection units, named \textit{X--ARAPUCA}. The detection of the prompt light pulse generated by the particle energy release in LAr will complement and boost the DUNE LAr Time Projection Chamber. It will improve the non--beam events tagging and enable at low energies the trigger and the calorimetry of the supernova neutrinos. The X--ARAPUCA is an assembly of several components. Its Photon Detection Efficiency~(PDE) depends on the design of the assembly, on the grade of the individual components and on their coupling. The X--ARAPUCA PDE is one of the leading parameters for the PDS sensitivity, that in turn determines the sensitivity of the DUNE for the detection of core-collapse supernova within the galaxy and for nucleon decay searches. In this work we present the final assessment of the absolute PDE of the FD1 X--ARAPUCA baseline design, measured in two laboratories with independent methods and setups. Preliminary results were reported in~\cite{Palomares_2022}. One hundred sixty units of these X--ARAPUCA devices have been deployed in the NP04 facility at the CERN Neutrino Platform, the 1:20 scale FD1 prototype, and will be operated during the year 2024. The assessed value of the PDE is a key parameter both in the NP04 and in the DUNE analysis and reconstruction studies.
\end{abstract}

\flushbottom

\section{Introduction} \label{sec:intro}

\begin{figure}[H]
    \centering
    \includegraphics[width=\linewidth]{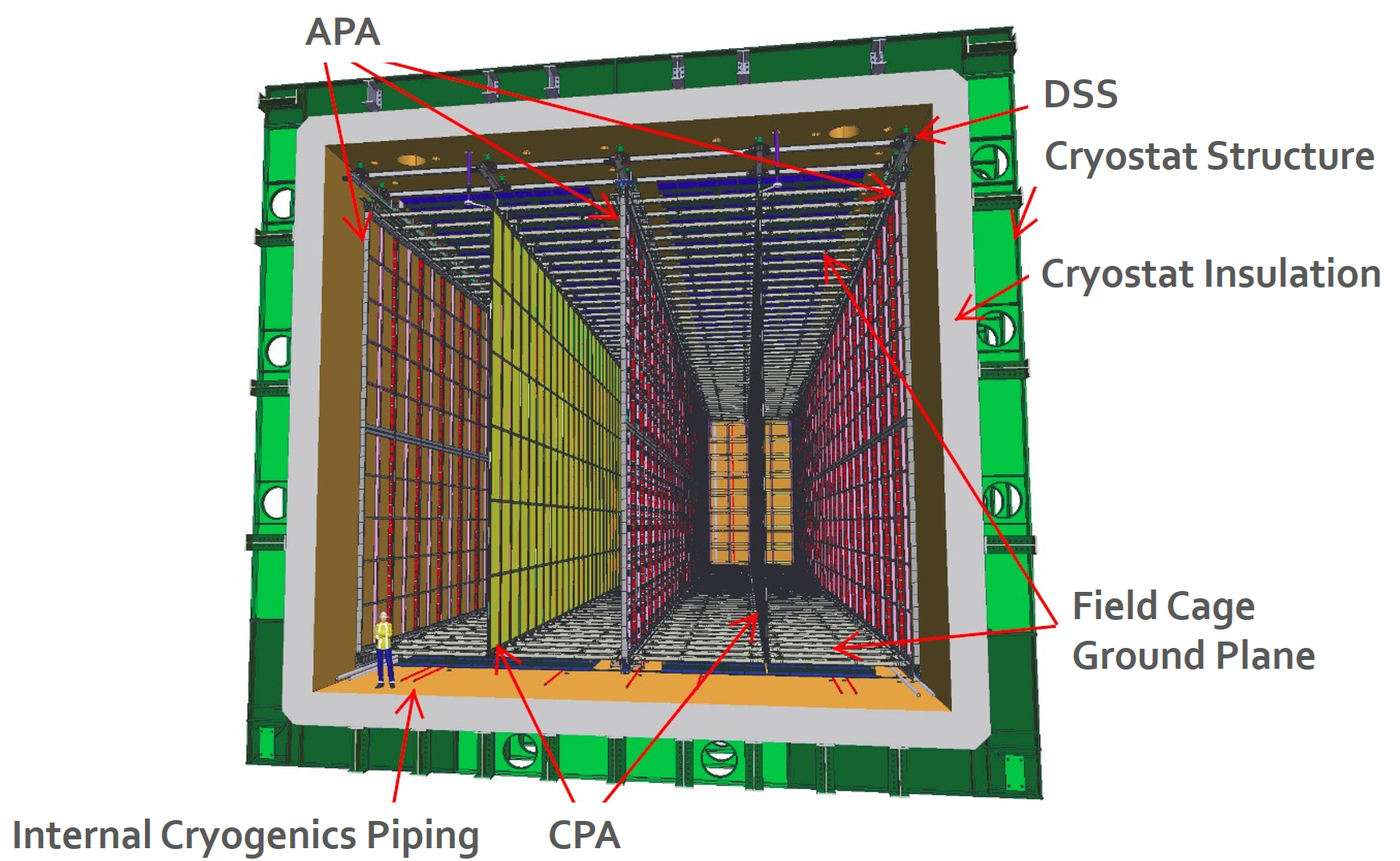}
    \caption{10 kt DUNE far detector module, showing the alternating 58.2~m long, 12.0~m high anode and cathode planes, as well as the field cage that surrounds the drift regions between the anode (APA) and cathode plane (CPA)~\cite{TDR1_1807.10334}.}
    \label{fig:DUNE-FD1}
\end{figure}

The Deep Underground Neutrino Experiment~(DUNE)~\cite{TDR1_1807.10334} is a dual--site experiment that aims to measure the neutrino oscillation parameters with a precision and sensitivity that will allow to test the CP violation in the leptonic sector and determine the neutrino mass ordering~\cite{TDR2_2002.03005}. It will also perform nucleon decay and beyond standard model searches and it will contribute to the detection of astrophysical neutrinos from the galaxy, the Sun and core--collapse supernova within the galaxy~\cite{10.1140-epjc-09166_SN}. The DUNE Far detector~(FD) will consist of four 17~kt LArTPC modules. This liquid argon~(LAr) technology will make possible to reconstruct neutrino interactions with image--like precision. The design of the four identically sized modules is sufficiently flexible for staging construction and evolving the LArTPC technology. The first FD Module~(FD1) will use the horizontal drift~(HD) technology, in which ionization charges drift horizontally in the LAr under the influence of an electric field towards a vertical anode, where they are read out. Four 3.5~m drift volumes are created between five alternating anode and cathode walls, each wall having dimensions of (58$\times$12)~m$^2$, and installed inside a cryostat. A schematic of the TPC is shown in Figure~\ref{fig:DUNE-FD1}.

LAr produces abundant VUV scintillation light, emitting 51000 photons/MeV when excited by minimum ionizing particles in the absence of a drift field~\cite{Doke_2002,Heindl_2010}. The particle's energy losses populate singlet and triplet states of Ar dimers~(Ar$_{2}^{\ast}$), that de--excite with characteristic times of 6~ns and about 1.6~$\mu$s respectively, emitting 127 nm photons.  
The high light yield and its efficient detection will enhance DUNE detector capabilities. To fulfill the supernovae neutrino program and for efficiently tagging nucleon decays, DUNE requires an average light yield of $>$20~photo--electrons~(PE)/MeV with a minimum $>$0.5~PE/MeV,  which corresponds to a collection efficiency of 1.3\% and 2.6\%, respectively~\cite{TDR3_2002.03008}.

The Photon Detector System of FD1 consists of light collector modules placed in the inactive space between the innermost wire planes of the anode planes.

These large--area light collectors convert incident scintillation photons into photons in the visible range that end up in photo--sensors. After investigating many light collector modules, DUNE has successfully reached a design for a light trap that can be deployed as either single--face or dual--face readout, the so--called X--ARAPUCA. In order to validate the DUNE technology, prototypes are being designed, built and tested at CERN, the ProtoDUNEs, LArTPC prototypes of 0.77kt total mass. ProtoDUNE--HD will test the final design of FD1 components at scale 1:1 in 2024. In particular, the X--ARAPUCAs will be tested for the first time there.

\section{The X--ARAPUCA device} \label{sec:XA}
\begin{figure}[htp]
	\centering
	\begin{subfigure}[b]{0.2\textwidth}
     	\centering
	\includegraphics[width=\linewidth]{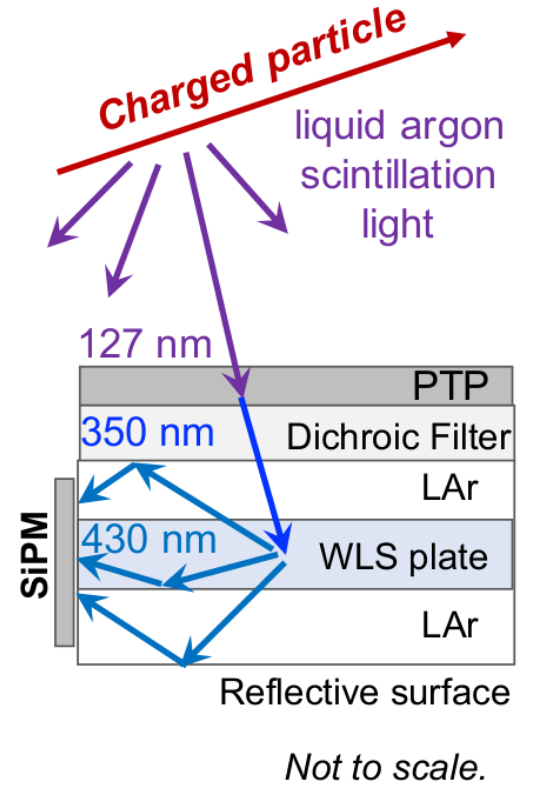}
	\caption{}
	\label{fig:working}
	\end{subfigure}%
	\begin{subfigure}[b]{0.28\textwidth}
    	\centering
	\includegraphics[width=\linewidth]{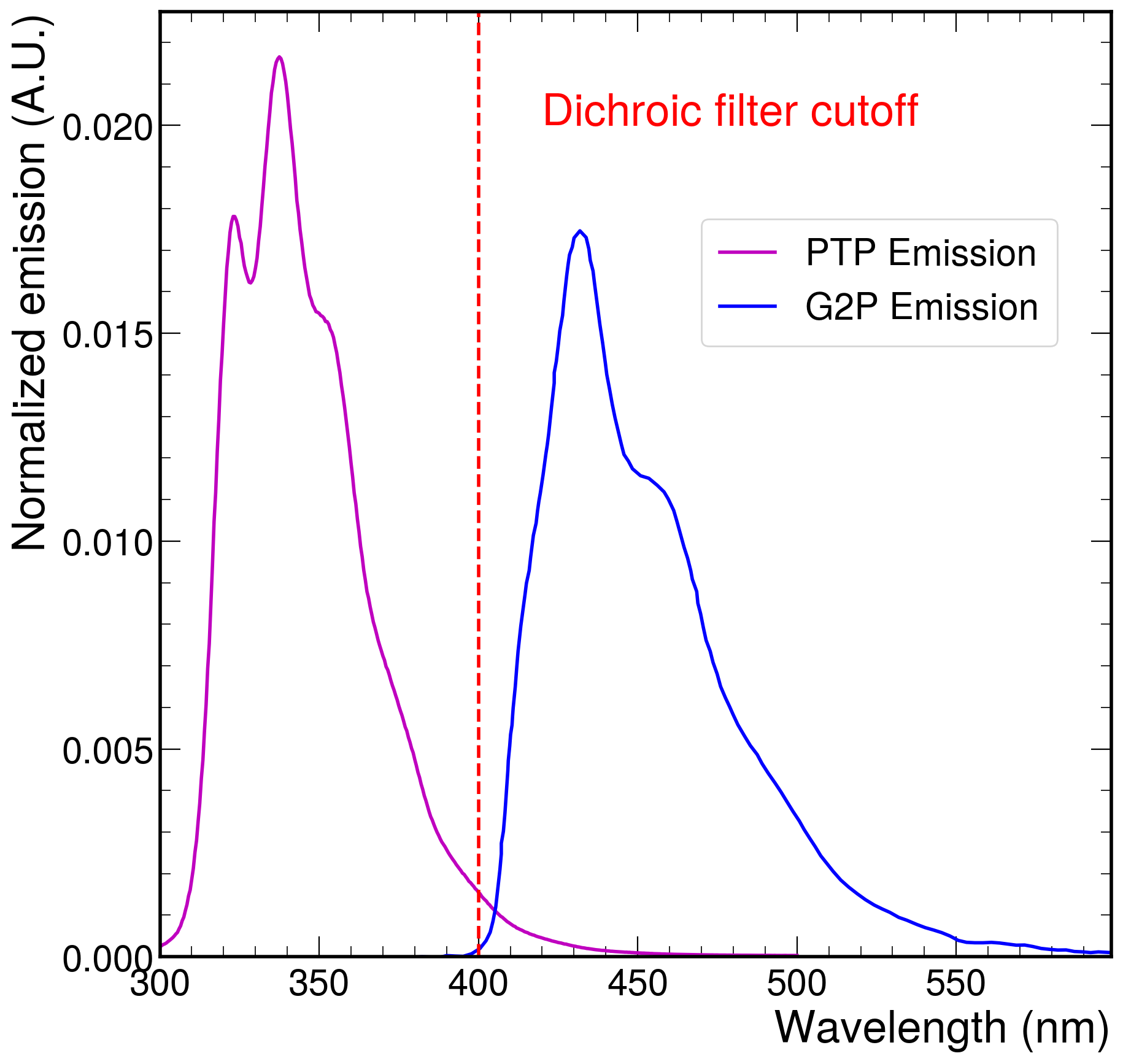}
	\caption{}
	\label{fig:emission}
	\end{subfigure}%

\medskip
 	\begin{subfigure}[b]{0.48\textwidth}
    	\centering
	\includegraphics[width=\linewidth]{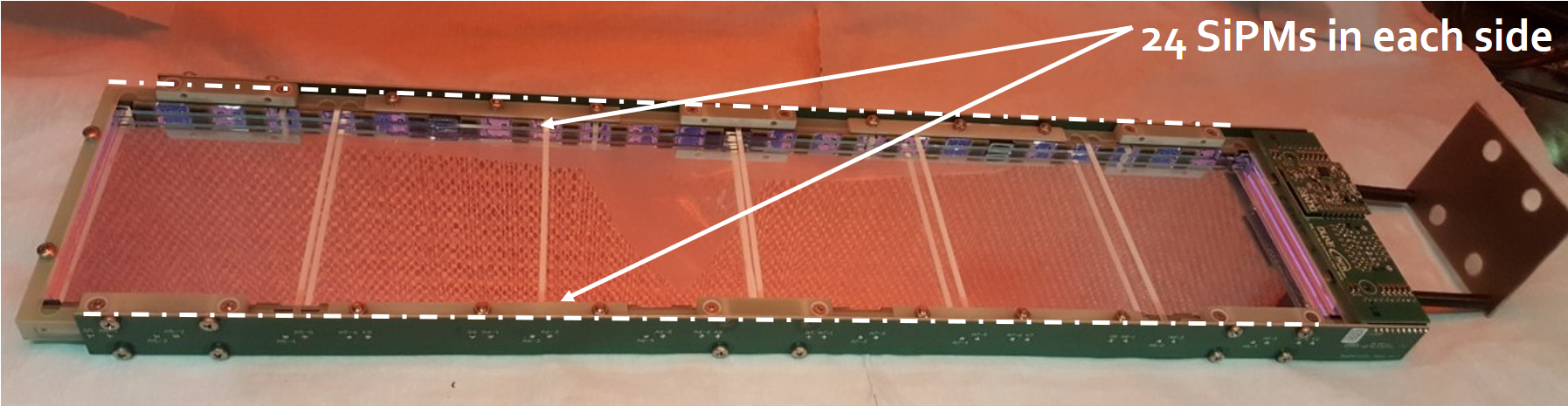}
	\caption{}
	\label{fig:sc_FBK}
	\end{subfigure}
	\caption{(a) Schematic of the XA working principle~\cite{TDR4_2002.031010v1}. (b) The dichroic filter cutoff (red dashed line), the pTP (purple line) and the G2P (blue line) emission spectra. (c) Image taken while mounting an FD1 XA of dimensions (50$\times$12)~cm$^2$.}
	\label{fig:XA}
\end{figure}

The X--ARAPUCA~(XA) is an improvement of the original concept of photon trapping inside a highly reflective box while using a wavelength shifting~(WLS) bar to increase the probability of collecting trapped photons onto a SiPM array. The ARAPUCA technology has been validated in the run--1 of the ProtoDUNE detector, that showed its superior performances in the photon detection efficiency compared to dip--coated shift lightguides (single or double)~\cite{DUNE:2020cqd}. In the XA the entrance window is a glass coated on the external side with a layer of p--Terphenyl~(pTP) ($\sim$500~$\mu$g/cm$^{2}$) to convert the incident 127 nm scintillation light into isotropically re--emitted 350 nm photons with an efficiency larger than 95\%. The re--emitted light reaches the WLS bar where it is further downshifted to the visible range~(430~nm), as illustrated in the schematic of Figure~\ref{fig:working}. The WLS re--emitted light can be trapped by total internal reflection~($\theta>56^{\circ}$) or escape the WLS bar~($\theta<56^{\circ}$) and be reflected by a dichroic filter and by a reflective surface. The filter is a multilayer thin-film coating deposited onto the inner side of the entrance window, whose cutoff is at 400 nm (Figure~\ref{fig:emission}). These trapped photons propagate towards the edges of the module where the Silicon Photo--Multiplier~(SiPM) photo--sensors are placed. Simulations of the X-ARAPUCA device~\cite{LPXAsim} showed the effectiveness of photon trapping inside the light guide and its dependence on factors such as light guide bulk transmittance, surface reflectiveness and optical coupling with the SiPMs. In DUNE, two different variants will be deployed: a double sided optical window variant, to equip the middle anode plane which collects scintillation photons from both the central drift volumes and a single sided variant to equip the two external anode planes. In this paper the single sided variant has been tested, in which the side opposite to the optical window is an opaque backplane lined inside with an extended specular reflector (ESR). Two different models for the WLS have been tested, the EJ-286 manufactured by Eljen Technology~\cite{Eljen} and the other custom designed and manufactured by Glass to Power (G2P)~\cite{G2P} in collaboration with INFN~\cite{INFN}. Figure~\ref{fig:emission} shows the emission spectra of both the G2P WLS and the primary wave length shifter. Earlier studies conducted at INFN showed the improved PDE performances of a smaller (or half-size) X-ARAPUCA device, embedding the G2P WLS than with the EJ286~\cite{PDE_XA_JINST}. 
The SiPM arrays are facing two of the sides of the WLS bar perpendicular to the entrance window. An image taken during the mounting process of an XA installed in Neutrino Platform 04 (NP04) is showed in Figure~\ref{fig:sc_FBK}. The SiPMs are evenly spaced and an ESR is placed between them on the left empty space on the mounting boards, so that photons reaching the edge of the module not hitting a SiPM are reflected back into the optical module. The SiPM active area is 6x6~mm while the light guide is 4~mm thick so that 1/3 of the area is exposed to the LAr allowing the detection of photons trapped by the dichroic filters.

\begin{figure}[htp]
	\centering
	\begin{subfigure}[b]{0.50\textwidth}
     	\centering
	\includegraphics[height=0.53\linewidth]{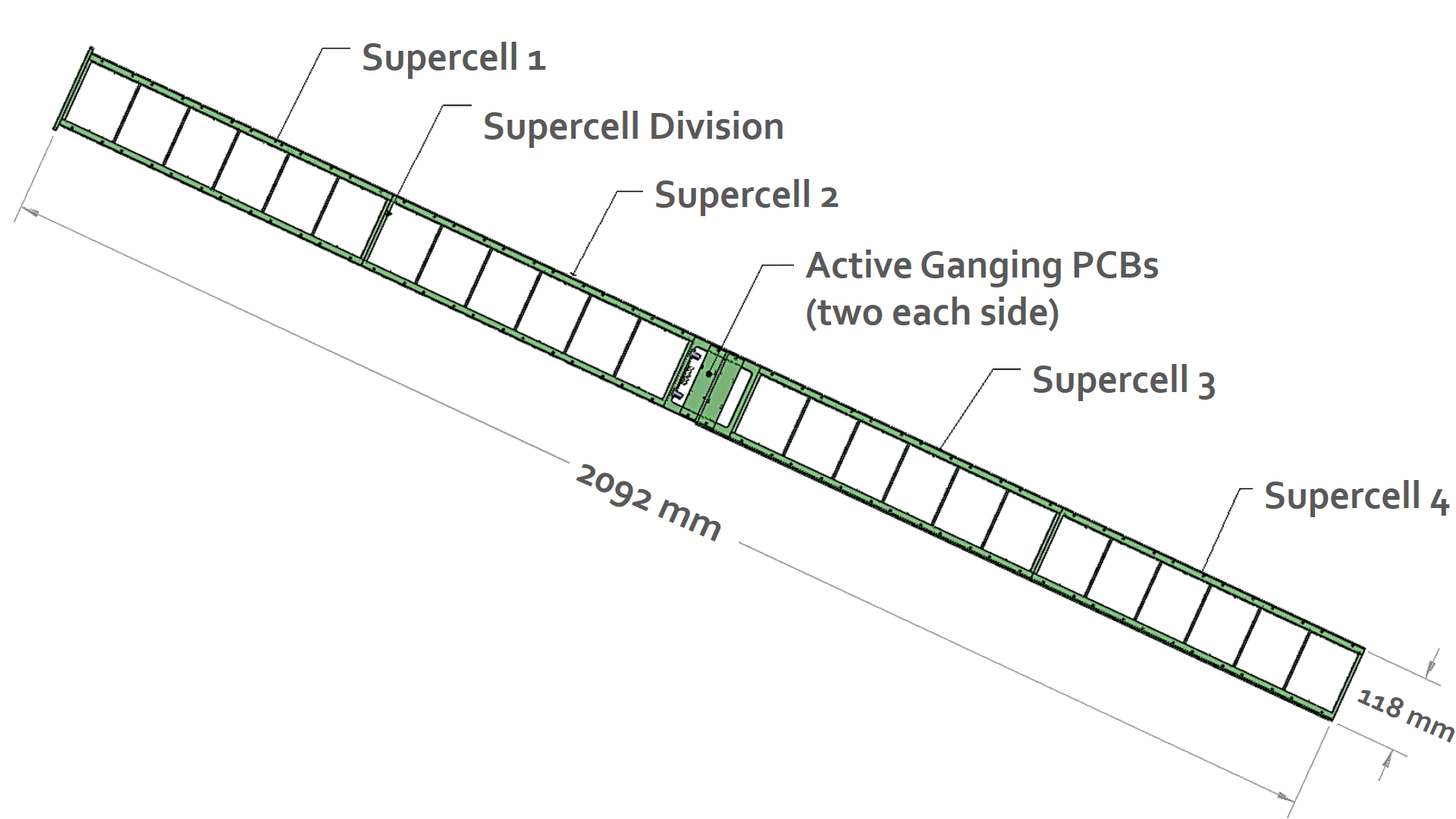}
	\caption{}
	\label{fig:module}
	\end{subfigure}%

	\begin{subfigure}[b]{0.50\textwidth}
    	\centering
	\includegraphics[height=0.58\linewidth]{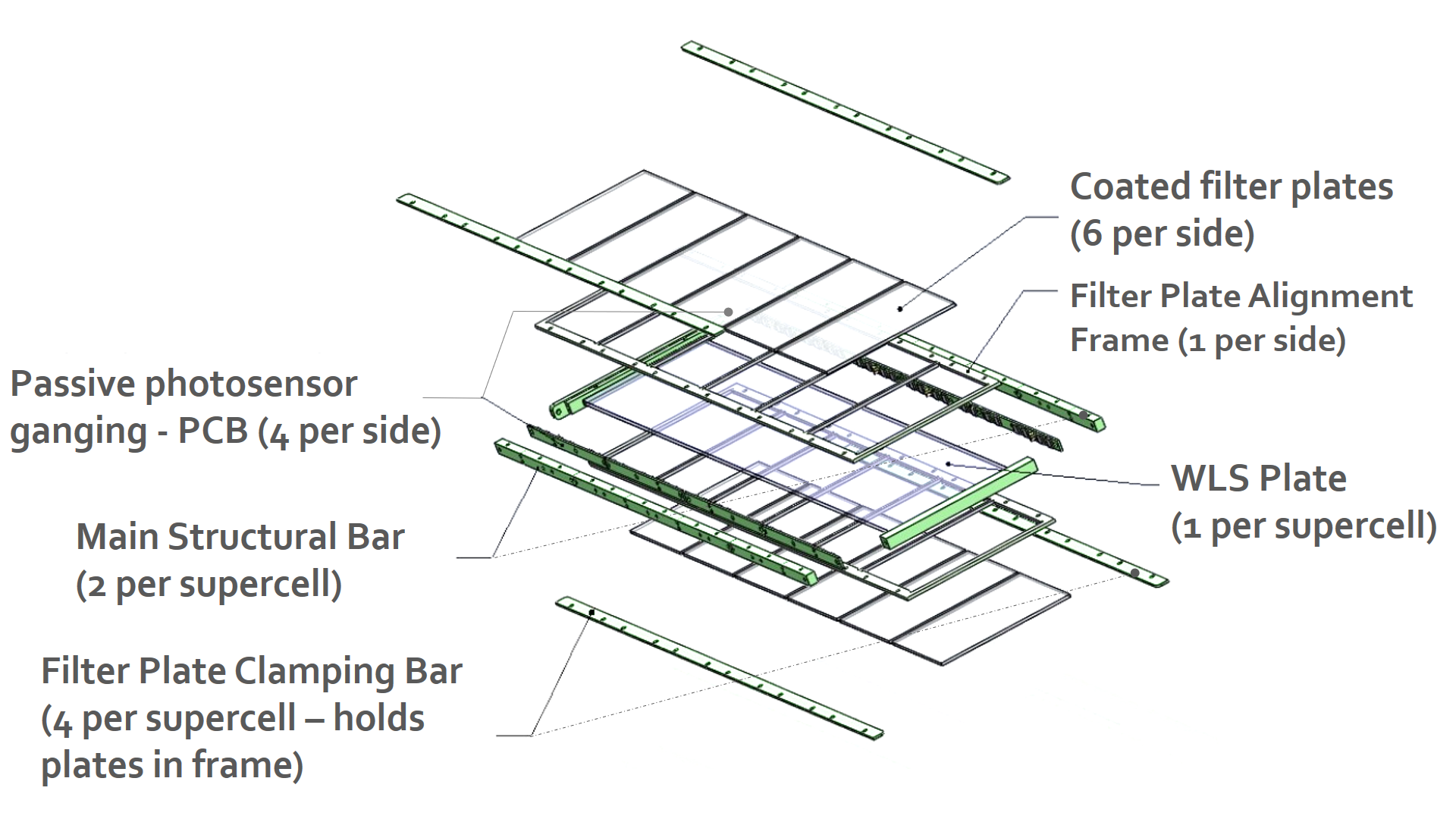}
	\caption{}
	\label{fig:components}
	\end{subfigure}
	\caption{(a) PDS module including 4 XA cells. (b) Schematic of the assembly process with the different components~\cite{TDR4_2002.031010v1}.}
	\label{fig:X-Arapuca}
\end{figure}

The module dimensions are 2~m long and 12~cm wide and it consists of 4 XA as shown in Figure~\ref{fig:module}, also known as \textit{Supercells}. In Figure~\ref{fig:components} we can see that each XA is composed of 6 dichroic filters, 1 WLS bar and 48 SiPMs grouped in 8 PCB boards (6 SiPMs each); the bias and signals are routed to the front end electronics by signal leading boards custom designed by INFN Sezione di Milano \cite{9459995}. 

Four configurations of these XAs depending on the SiPMs and WLS bar type will be tested in ProtoDUNE. Two different models of SiPMs, from Fondazione Bruno Kessler~(FBK) and Hamamatsu Photonics K.K~(HPK), will be employed. The particular models are: FBK Triple--Trench~(TT)~\cite{FBK}, which pixel size is about 50~$\mu$m, and HPK (S13360-6075HS-HRQ)
75~$\mu$m High Quenching Resistance~(HQR)~\cite{HPK_SiPM} both with a total effective area of about 36~mm$^{2}$ and specifically designed for being used at cryogenics temperatures~(CT). ProtoDUNE--HD also makes use of two different WLS bars, the EJ--286PS and the bar custom designed by G2P. For all the four configurations OPTO~\cite{opto_br} dichroic filters were used. In Table~\ref{tab:xa_combinations} we summarize the four combinations installed in ProtoDUNE--HD that have been characterized in this work.

\begin{table}[htp]
\centering
\resizebox{\columnwidth}{!}{%
	\begin{tabular}{llll}
	\hline
 	& SiPMs & WLS bar & Testing Site \\ \hline \hline
	(A) & {FBK TT} & EJ--286PS--1 & CIEMAT + MiB \\
	(B) & {FBK TT} & G2P--FB165A & MiB \\
	(C) & {HPK 75HQR} & EJ--286PS--1 & CIEMAT \\
	(D) & {HPK 75HQR} & G2P--FB165A  & CIEMAT + MiB \\ \hline
	\end{tabular}%
}
	\caption{The four XA configurations installed in ProtoDUNE--HD.}
	\label{tab:xa_combinations}
\end{table}

\section{Methodology and Instrumentation} \label{sec:measurement} 
Quantifying the absolute efficiency is fundamental for DUNE, as this parameter is needed to fully characterize the PDS.
The PDE must be assesed at the operative conditions, at CT and with 127 nm photons. For this reason, we submerged the XA in LAr together with a low--activity electro--deposited $^{241}$Am alpha source. Two experimental setups have been established to carry out this measurement, at CIEMAT (Madrid, Spain) and at Milano--Bicocca University (Milan, Italy). To measure the absolute efficiency of the XA, the number of photons arriving at its surface need to be known. Two different methods have been considered: comparing the amount of light collected by the XA with the light detected by a calibrated photosensor, and estimating the light in the XA from the $\alpha-$source energy and the known number of scintillation photons per MeV produced in LAr once the solid angle sustained by the source is determined. Both setups liquefy high purity argon gas~(GAr) inside a cryogenic vessel.

\begin{figure}[htp]
	\centering
	\includegraphics[width=\linewidth]{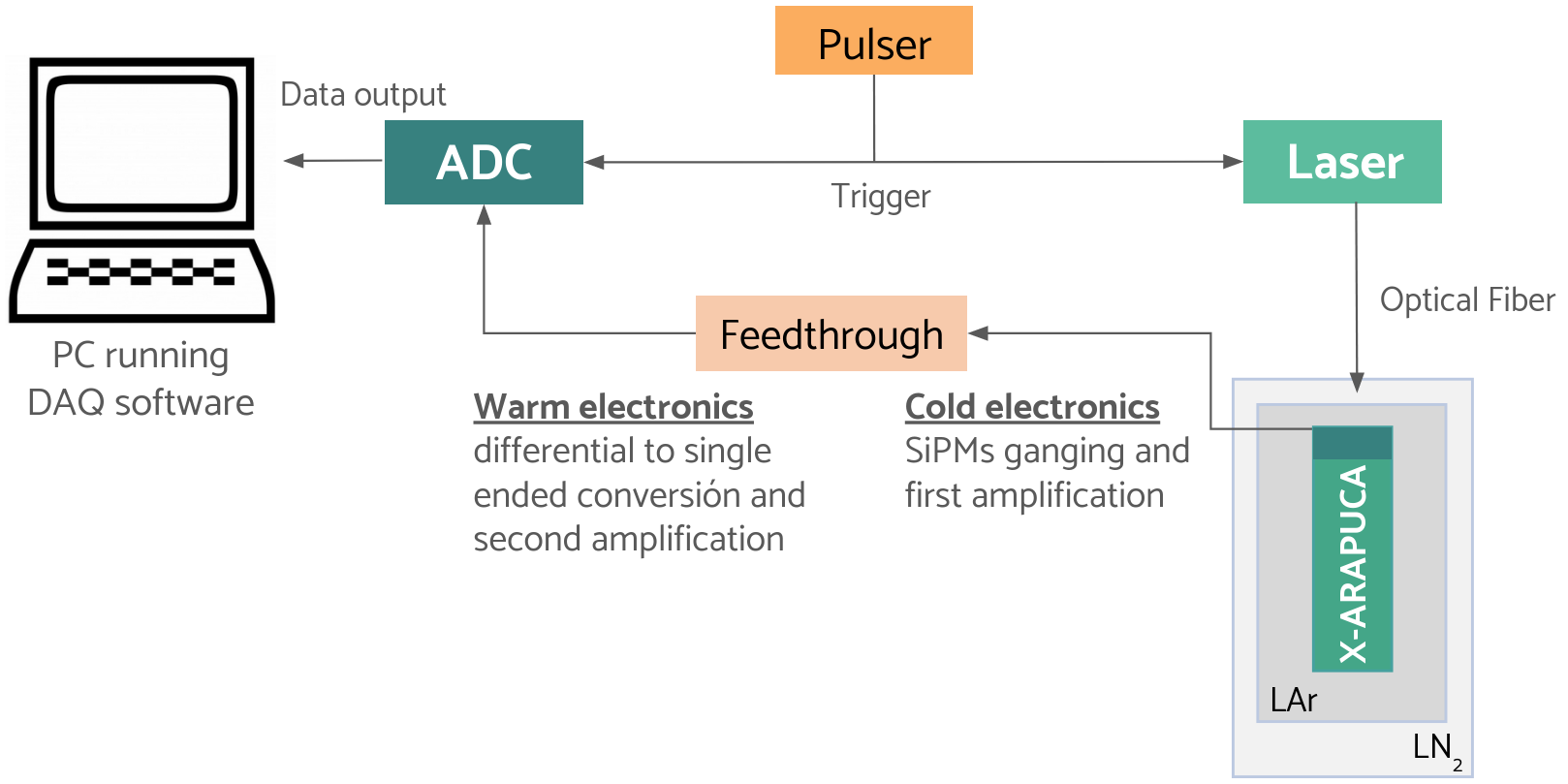}
	\caption{Schematic of the data acquisition system.}
	\label{fig:DAQ}
\end{figure}

The data processing and acquisition~(DAQ) system configuration, involving the software and the electronics, is similar in both setups and a diagram is shown in Figure~\ref{fig:DAQ}.

The 48 SiPMs are passively grouped at the input of a transimpedance amplifier located at the top of the XA, the so--called \textit{cold--amplifier}. The amplifier is based on a SiGe bipolar transistor (Infineon BFP640) followed by a fully differential operational amplifier (Texas Instruments THS4531), and is designed for low noise at low power, giving a voltage white noise density of 0.37 nV/$\surd$Hz at 2.4 mW per channel. To reduce the noise that may be induced in signal cables, the output of the cold--amplifier is differential. At room temperature, the so--called \textit{warm--electronics} converts the differential signal to single ended and introduces a second amplification factor. The readout scheme replicates the one planned for the DUNE FD1, where the differential to single ended conversion is performed in AC using a transformer, as described in~\cite{brizzolari_2022}. This gives an undershoot on the tail of the signals, which will need to be considered in the analysis.
Both the cold and the warm electronics are the same as in the MiB and CIEMAT setups, while the adopted digitizer differs, as reported in the following sections~\ref{sec:ciemat_setup} and~\ref{sec:mib_setup}.

\subsection{CIEMAT setup}\label{sec:ciemat_setup}
\begin{figure}[htp]
	\centering
     \begin{subfigure}[b]{0.37\textwidth}
         \centering
        \includegraphics[width=\linewidth]{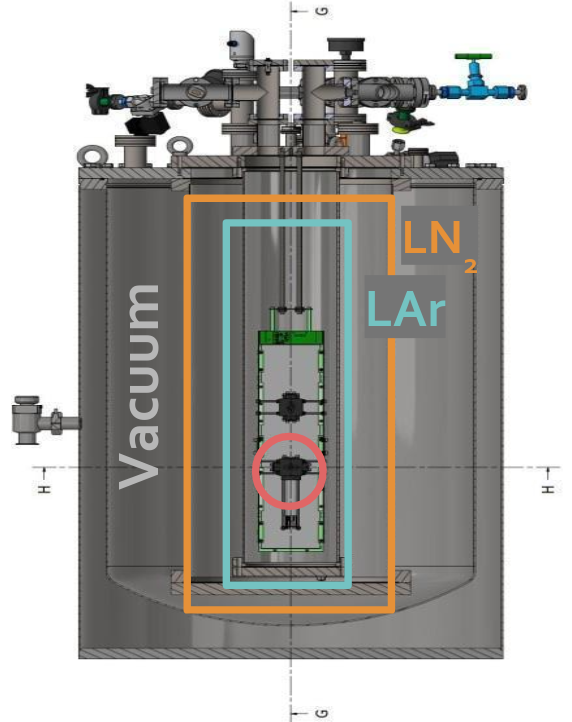}
        \caption{}
        \label{fig:vessel_LAr}
    \end{subfigure}%
    
    \begin{subfigure}[b]{0.47\textwidth}
        \centering
        \includegraphics[width=\linewidth]{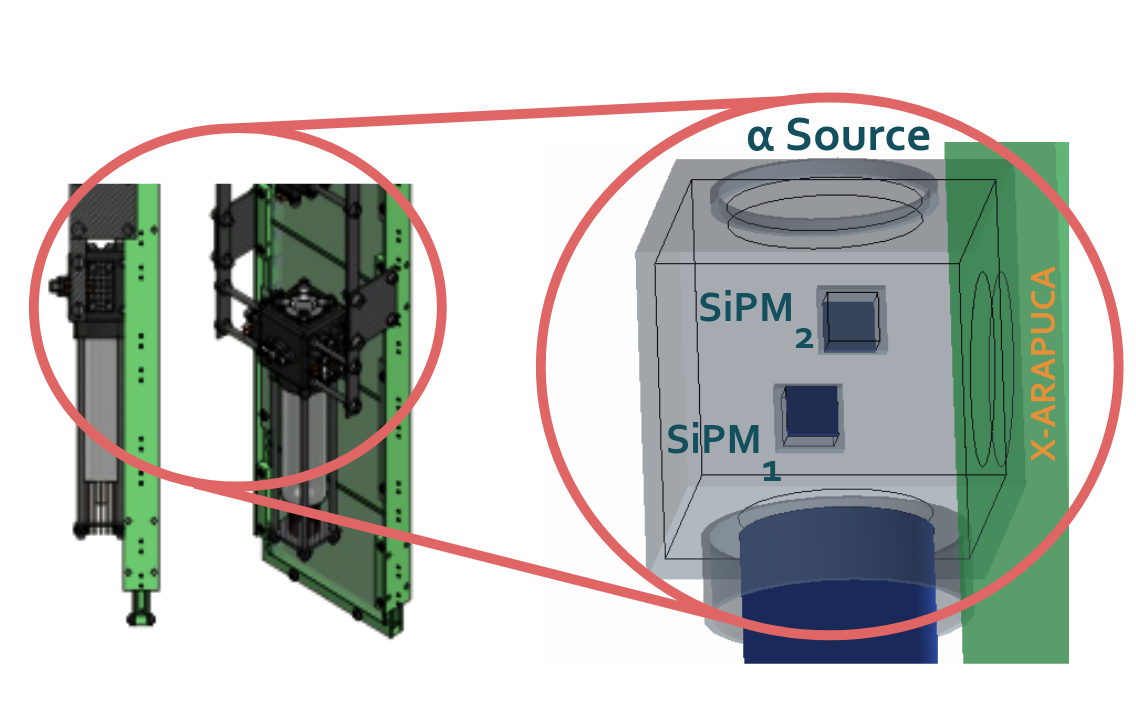}
        \caption{}
        \label{fig:box_LAr}
    \end{subfigure}
	\caption{Setup scheme used for obtaining the absolute efficiency of the XA. (a) Cryogenic vessel with its concentric volumes where GAr is liquified. (b) Diagram of the black box holding the sensors together with the $^{241}$Am source.}
	\label{fig:setup_ciemat}
\end{figure}

The CIEMAT neutrino group made use of a 300~l cryogenic vessel with different concentric volumes, whose schematic is shown in Figure~\ref{fig:setup_ciemat}. The larger and external one~(100~l) is filled with liquid nitrogen~(LN$_2$) and the smaller one~(18~l) contains the GAr and it is where the XA is located. In this 18~l container, the GAr is liquefied by thermal contact with the LN$_2$ of the surrounding volume. This is achieved by controlling pressure parameters and regulating the temperature values necessary to carry out the liquefaction. The system is designed to perform the automatic filling of the 100~l vessel with LN$_2$ from a 400 l tank which is at an over--pressure of 4~bar. At the end, GAr grade 6.0 is liquefied with LN$_2$ at 2.7~bar. To reduce contamination from material outgassing, we perform successive vacuum cycles in the vessel before introducing the optical and electrical components that will be used to perform the measurements. A last vacuum cycle is done once all elements are in place. Each XA configuration is tested in data taking campaigns lasting 3--4 days.

The XA is introduced in the inner vessel together with two reference HPK VUV4 SiPMs (S13370--6075CN)~\cite{HPK_SiPM} and a Photo--Multiplier Tube~(PMT) (R6836--Y00)~\cite{HPK_PMT} as complementary photo--sensors. The VUV4 SiPMs are designed to have a high sensitivity for VUV light and stable performance at CT, making possible the detection of LAr scintillation light. The PMT is not specifically designed to measure at CT, although it is also sensitive to VUV light, allowing us to study the scintillation light profile easily and monitor the LAr purity.  

Together with the reference detectors a $^{241}$Am source is held with an opaque box as shown in Figure~\ref{fig:setup_ciemat}. It emits $\alpha$-particles with 5.485~MeV (84.45\%) and 5.443~MeV~(13.23\%) energies with an activity of (54.53~$\pm$~0.82)~Bq~\cite{2023.110913_Validation241AmLiquifiedGasDetectorsCT}. The particles deposit their energy inside the 4~cm sized black box, and the produced photons reach the XA through a hole ($\varnothing$~=~23~mm). We ensure that no other photons are being detected by covering the rest of the XA with a black sheet. On the other faces of the box, we place the two VUV4 SiPMs, the PMT and a diffuser connected to a fiber. The optical fiber provides light from LEDs and lasers to check the response of the system. In Table~\ref{tab:dimensions} we summarize the detectors' dimensions and positions in the box.

\begin{table}[htp]
\centering
\resizebox{\columnwidth}{!}{%
	\begin{tabular}{lcc}
	\hline
                                     	& XA             	& Ref. SiPM \\ \hline \hline
	Effective area (mm$^2$)          	& 415.47         	& 36.00   \\
	Distance to the source (mm) & 29.0  $\pm$  0.7   & 26.9  $\pm$  0.3 \\ \hline
	\end{tabular}%
}
	\caption{Relative distances from the $\alpha$--source to the sensors and sizes of the box's holes.}
	\label{tab:dimensions}
\end{table}

The VUV4 SiPMs were calibrated at room temperature by the manufacturer; however, several studies~\cite{2202.02977_SiPMVUV4_PDE_CT+127nm,1910.06438_Reflectivity+PDE_HPK_VUV4SiPMs_LXe,1912.01841_ReflectanceVUVS13370-6075CN} have shown a decrease of about 50\% in the PDE at CT and a dependence with the incident angle. Considering the results of these studies and the measurements carried out in CIEMAT labs~\cite{HPK_VUV4_PDE_CIEMAT}with these sensors exposed to VUV light at different angles, we will assume a PDE of (11.17~$\pm$~1.3)\% at 127~nm and, 87~K and 4~V overvoltage~(OV) for the model S13370--6075CN of VUV4 SiPMs from HPK. The cross--talk probability computed in~\cite{HPK_VUV4_PDE_CIEMAT} for the reference SiPMs at CT is $\rm P_{XT} = (14.84~\pm~0.24)\%$ which is in perfect agreement with the result presented in~\cite{2202.02977_SiPMVUV4_PDE_CT+127nm}.

The XA signal provided by the warm--electronics is digitized by the ADC (model CAEN DT5725S~\cite{CAEN}). The signals of the two reference SiPMs are amplified and digitized by the same CAEN module. The trigger is done at ADC level and is provided by the signal in coincidence of the two reference SiPMs. The final output consists of 20 $\rm \mu s$ waveforms with 4~ns sampling. A pre--trigger of 2 $\rm \mu s$ allows the determination of the baseline on a event--by--event basis. For calibration runs, a pulse generator provides an external trigger and synchronously pulses the light source (laser or LED). Three--$\rm \mu s$ waveforms are recorded in the calibration runs with the same time sampling.

\subsection{INFN Milano Bicocca setup}\label{sec:mib_setup}

The INFN Milano--Bicocca~(MiB) setup is an extension of the one used for the PDE measurements of the XA device adopted by the SBND project~\cite{1702.00990_ProgramSBND}. 
The work~\cite{PDE_XA_JINST} allowed us to precisely assess the superior performance of the SBND-XA device equipped with the custom produced PMMA based WLS~\cite{G2P}, that is now adopted as the baseline component for both the DUNE FD1 and FD2 Photon Detection systems. The setup, the procedures and the method are described in detail in~\cite{PDE_XA_JINST}.

\begin{figure}[htp]
	\centering
	\begin{subfigure}[b]{0.40\textwidth}
     	\centering
	\includegraphics[height=\linewidth]{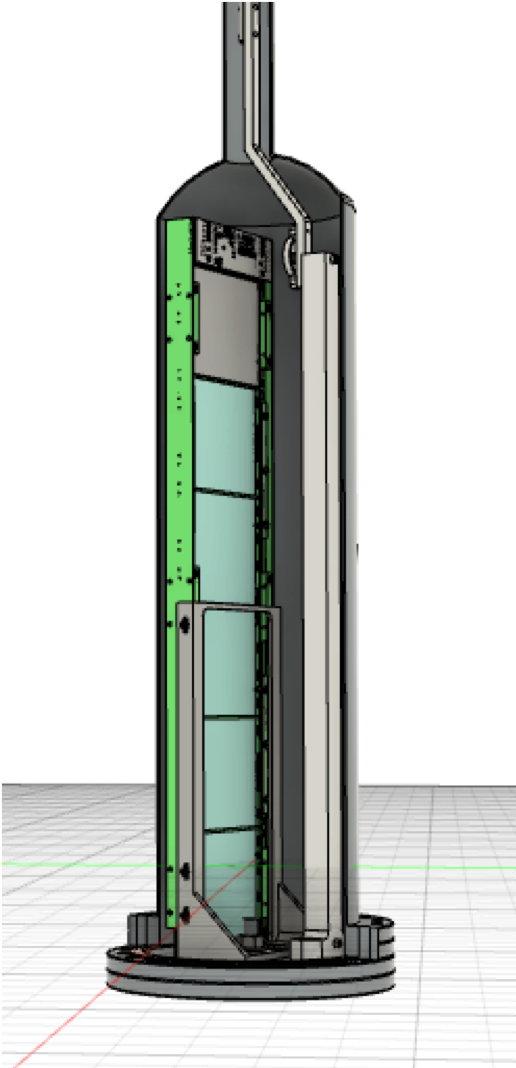}
	\caption{}
	\label{fig:cad}
	\end{subfigure}%
 
	\begin{subfigure}[b]{0.37\textwidth}
    	\centering
	\includegraphics[height=\linewidth]{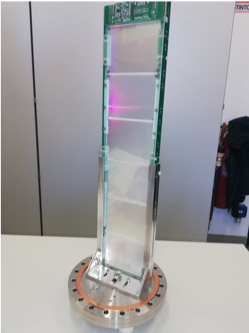}
	\caption{}
	\label{fig:onflange}
	\end{subfigure}
	\caption{(a) The scheme of the INFN--MiB setup for the longitudinal scanning and PDE determination of the XA. (b) Picture  before closing the chamber.}
	\label{fig:setup_mib}
\end{figure}

Figure~\ref{fig:setup_mib} shows the XA located at the center of the stainless steel $\sim$25~l cylindrical chamber of 250~mm diameter  and 550~mm height. The closed chamber is located in an open 70~l dewar. The chamber is first outgassed down to $O({10^{-4})}$~mbar and then connected to a bottle of GAr grade 6.0. The open dewar is filled with LAr and the GAr liquefaction process starts inside the chamber. Both the GAr flow and its liquefaction rate in the chamber are sustained by the regulation of the bottle pressure reducer. The exposed $^{241}\rm{Am}$ $\alpha$--source (3.7~kBq) is mounted on the tip of a magnetic manipulator (rototraslator) and allowed to slide on a vertical rail, facing the XA at the distance of ($55\pm1$)~mm. This allows to scan the PDE of the XA along its z--axis and to monitor with high precision  the LAr level inside the chamber during the whole filling process, as the alpha light pulse amplitude greatly increases when the source is in LAr.

The target operational condition is reached when the XA and the front end readout circuit are submerged in LAr. Due to the lack of GAr, in few runs   we couldn't reach the top of the device, hence only the data from the scanning positions fully submerged by LAr are considered in the PDE data analysis.

%
The digitizer (CAEN DT5725 250~MS/sec 14~bits) is self--triggered by setting a threshold, that provides a  trigger rate of about 1~kHz for the $\alpha$--particles and 100~Hz for the muons runs respectively.

\subsection{Monte Carlo simulations}\label{sec:MC}
To properly assess the PDE of the XA prototype, a dedicated GEANT4~\cite{GEANT4} Monte Carlo~(MC) simulation was developed for each setup. For both of them, the scintillation photons generated by the alpha particle energy loss in LAr are emitted uniformly and isotropically, and the MC provides the number of photons reaching the XA acceptance window. For the CIEMAT setup the reference photosensors solid angles are also determined by the simulations and for both MiB and CIEMAT the geometrical acceptances uncertainties are computed by varying both the position and dimensions of the setup elements within their precision errors. 

\begin{figure}[htp]
	\centering
	\begin{subfigure}[b]{0.5\textwidth}
     	\centering
	\includegraphics[height=0.45\textwidth]{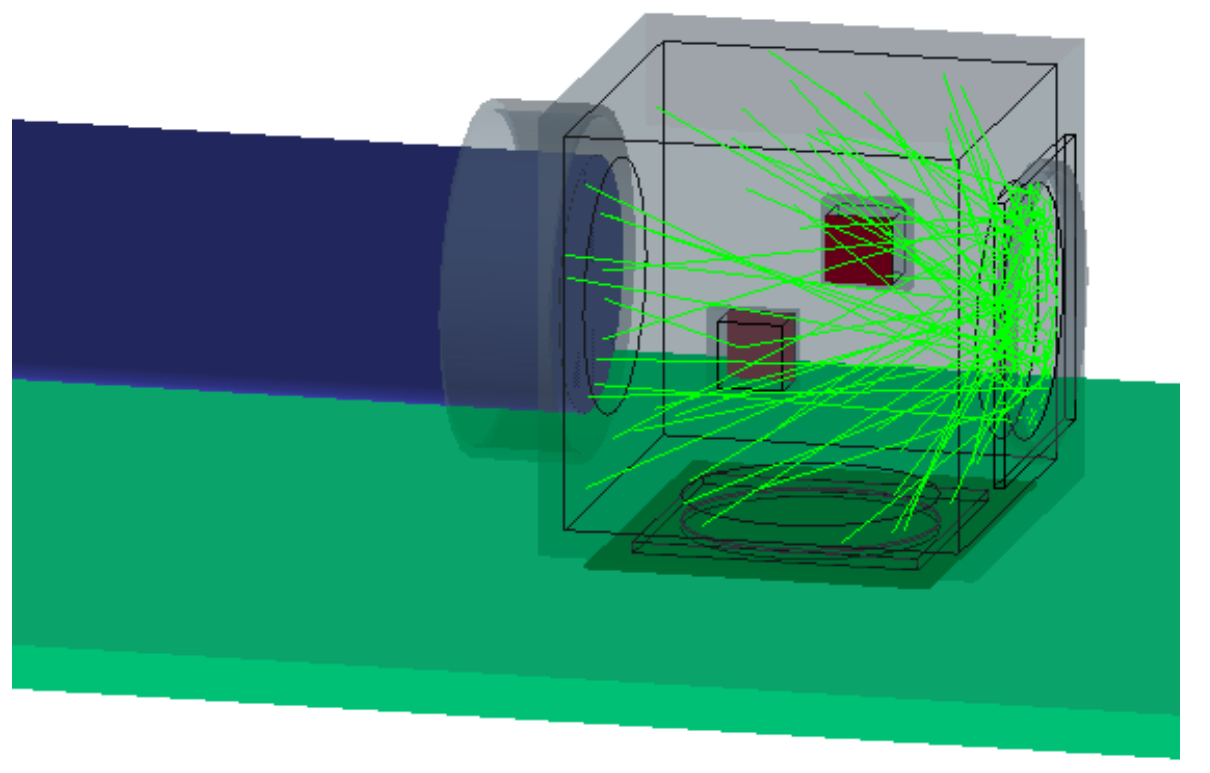}
	\caption{}
	\label{fig:CIEMAT_MC}
	\end{subfigure}%

	\begin{subfigure}[b]{0.5\textwidth}
    	\centering
	\includegraphics[height=0.23\linewidth]{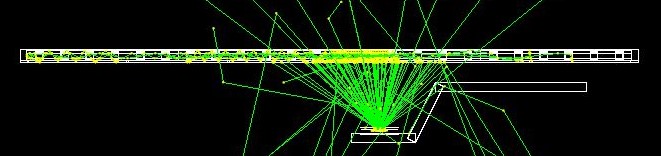}
	\caption{}
	\label{fig:MiB_MC}
	\end{subfigure}
	\caption{Example of events generated for (a) CIEMAT and (b) Milano--Bicocca setups.}
	\label{fig:MC}
\end{figure}

Figure~\ref{fig:CIEMAT_MC} shows the simulation of the CIEMAT setup with the dimensions presented in Table~\ref{tab:dimensions}.  The two reference VUV4 SiPMs (red), the PMT (blue) and the XA (green) are configured as sensitive materials to retrieve the number of detected photons depending on their position with respect to the source. The surrounding black box is designed with a black plastic material that fully absorbs the photons. Each event of the alpha source is simulated to have the number of photons in a random position within the sensitive area. We have determined a systematic error due to the geometrical acceptance uncertainty of 10.8\%.

Figure~\ref{fig:MiB_MC} shows a side view of the XA long edge side of the Milano Bicocca MC geometry. About 500 photons are drawn for visualization purposes. The $^{241}$Am deposited surface exposed to LAr is embedded in a source holder whose shape and size is included in the MC model to determine the effective light cone. The geometrical acceptance is then driven by the device-to-source distance, that is measured $(5.5 \pm 0.1)$ cm, along the whole rail length. The uncertainty on the geometrical acceptance accounts for $\sim7\%$ and represents the major systematic error for the MiB method. The LAr optical properties are simulated too but for optical path in LAr of ($O(10$ cm$)$) both the absorbance and the Rayleigh scattering are negligible.

\section{Data Analysis} \label{sec:analysis_CIEMAT}

\subsection{Calibration} \label{sec:ana_cal}

To measure the PDE, the light pulse generated in LAr by $\rm ^{241}Am$ alpha particles and collected by the photosensors must be calibrated by the integrated charge of the single photon electron, i.e. by the gain factor of the SiPMs.  The gain depends on the operative temperature and bias voltage, and can be affected by the {\it fatigue effect}, like for the photo--multiplier tubes~\cite{1985AJ_PMTFatigue}. In both the CIEMAT and the MiB setup the calibrations are performed with low intensity blue light emitting sources.

\begin{figure}[htp]    
	\centering
	\begin{subfigure}[b]{0.5\textwidth}
     	\centering
	\includegraphics[height=0.74\linewidth]{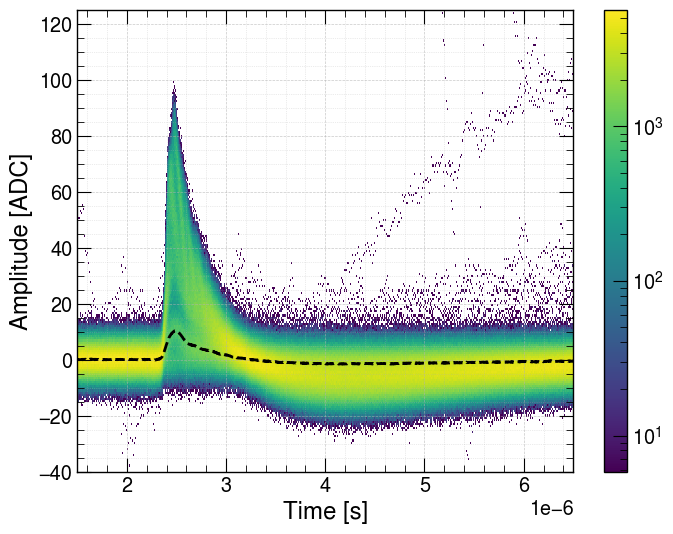}
	\caption{}
	\label{fig:persistence}
	\end{subfigure}%

	\begin{subfigure}[b]{0.5\textwidth}
    	\centering
	\includegraphics[height=0.66\linewidth]{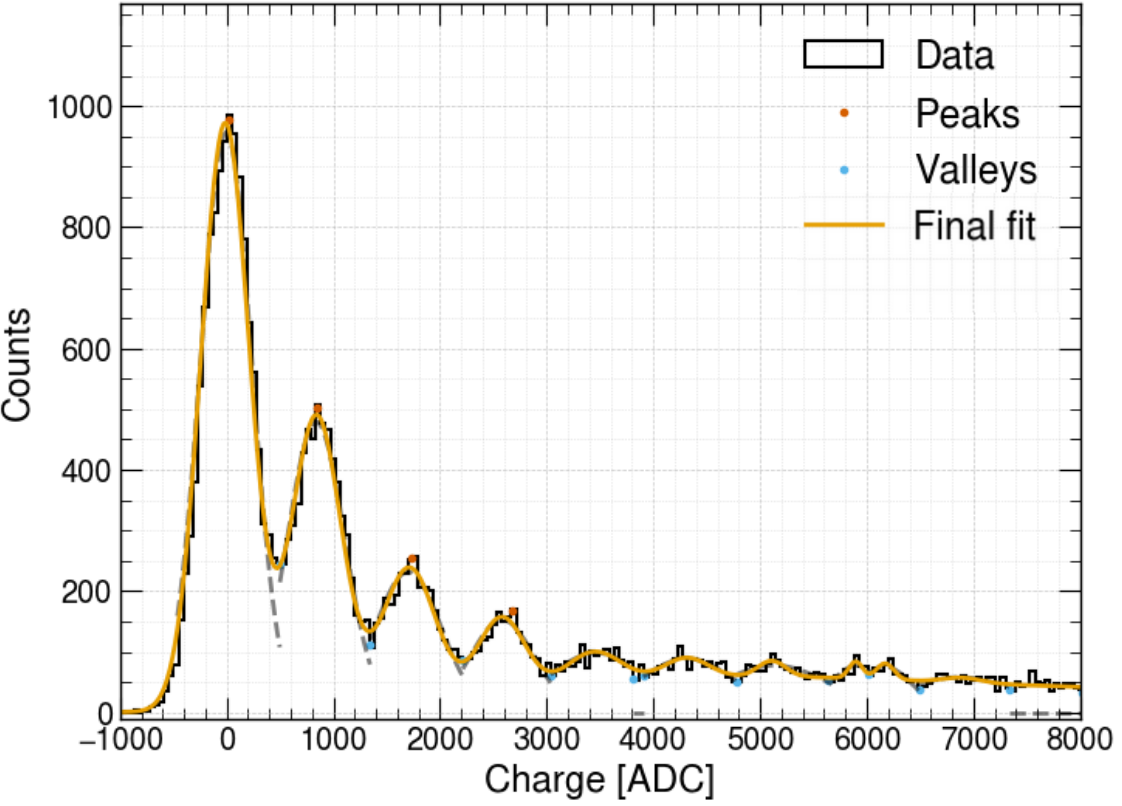}
	\caption{}
	\label{fig:calibration_hist}
	\end{subfigure}
	\caption{(a) Persistence histogram of selected waveforms together with the average waveform of one single PE. (b) Charge histogram of selected peaks, the gain is defined by the distance between the first and the second peak.}
	\label{fig:calibration}
\end{figure}

Typical calibration waveforms for the XA are displayed in Figure~\ref{fig:persistence}. The gain is determined from the integrated charge distribution, as shown as an example in Figure~\ref{fig:calibration_hist} and defined as $\rm Gain=(\mu_{2}-\mu_{1})$ where $\mu_n$ is the mean value of the Gaussian corresponding to $n$ photoelectrons. The good signal--to--noise ratio~(SNR) allows the identification of the peaks corresponding to 1, 2, 3 to N PE and then the gain determination.

The SNR qualifies the capability to detect a single PE~(SPE) over the system noise: it depends both on the XA electronics and on the setup related disturbances. We define SNR $= \frac{\mu_1 - \mu_0}{\sqrt{\sigma^2_0 + \sigma^2_1}}$ from the integrated charge distribution where $\sigma_n$ is the Gaussian width of the nth peak (0 is the baseline noise peak, 1 is the 1PE peak).
For each photosensor type, Table~\ref{tab:snr} reports the SNR measured for three overvoltage bias values. In all cases a SNR~$>2$ was measured, hence in all the measurements the SPE detection capability is verified.
\begin{table}[htp]
	\begin{subtable}[t]{.24\textwidth}
    	\centering
    	\begin{tabular}{ccc}
        	\hline
        	\multicolumn{1}{c}{OV} & PDE & \multicolumn{1}{c}{SNR} \\ \hline \hline
        	3.5 & 40 & 2.40 $\pm$ 0.08 \\
        	4.5 & 45 & 3.42 $\pm$ 0.07 \\
        	7.0 & 50 & 3.76 $\pm$ 0.05 \\ \hline
    	\end{tabular}
    	\caption{FBK TT}
	\end{subtable}%
	\begin{subtable}[t]{.24\textwidth}
    	\centering
    	\begin{tabular}{ccc}
        	\hline
        	\multicolumn{1}{c}{OV} & PDE &\multicolumn{1}{c}{SNR} \\ \hline \hline
        	2.0 & 40 & 2.90 $\pm$ 0.03 \\
        	2.5 & 45 & 3.55 $\pm$ 0.02 \\
        	3.0 & 50 & 4.24 $\pm$ 0.02 \\ \hline
    	\end{tabular}
    	\caption{HPK HQR75}
	\end{subtable}
\caption{Experimentally measured SNR of the XA SiPMs at a given overvoltage (OV) and corresponding Photon Detection Efficiency (PDE).}
\label{tab:snr}
\end{table}
 
\begin{figure}[htp]    
 \centering
	\includegraphics[width=\linewidth]{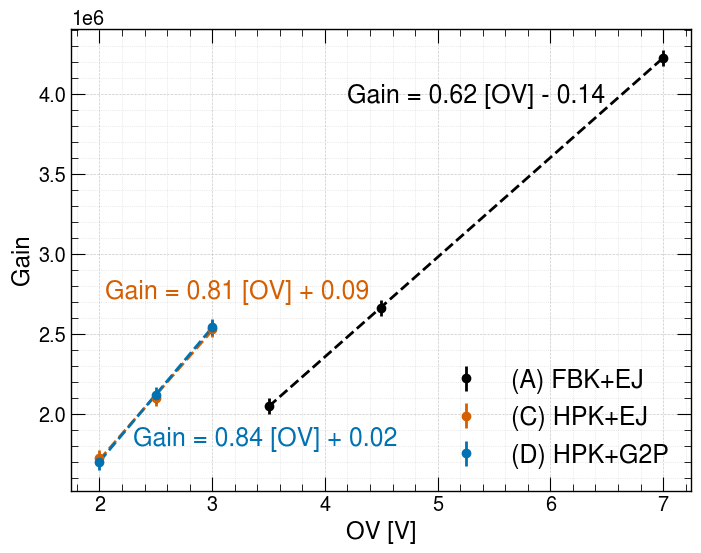}
	\caption{Gain versus bias voltage. Dashed lines represent linear ﬁts of the experimental points. Unit conversion to get adimensional gain values was made using the amplifier's gain and the electron charge.}
	\label{fig:gains_xa}
\end{figure}

Figure~\ref{fig:gains_xa} features the measured gain versus over--voltage of XAs equipped with different SiPM models. The gain is a characteristic of the SiPM and independent of the rest of the XA elements.

Dedicated cross--talk studies have been performed by the PDS Consortium for the two SiPMs models deployed in the XAs. The measured cross-talk probabilities~($\rm P_{XT}$, probability to have a second pixel activation after a true photo electron) are presented in Table~\ref{tab:x-talk_probs} for three OV bias values and are used to compute the correction for the efficiency.
\begin{table}[htp]
	\begin{subtable}[t]{.24\textwidth}
    	\centering
    	\begin{tabular}{ccc}
        	\hline
        	\multicolumn{1}{c}{OV} & PDE & \multicolumn{1}{c}{$\rm P_{XT}$ (\%)} \\ \hline \hline
        	3.5 & 40 & 12.68 $\pm$ 0.27 \\
        	4.5 & 45 & 16.05 $\pm$ 0.32 \\
        	7.0 & 50 & 32.47 $\pm$ 0.47 \\ \hline
    	\end{tabular}
    	\caption{FBK TT}
	\end{subtable}%
	\begin{subtable}[t]{.24\textwidth}
    	\centering
    	\begin{tabular}{ccc}
        	\hline
        	\multicolumn{1}{c}{OV} & PDE &\multicolumn{1}{c}{$\rm P_{XT}$ (\%)} \\ \hline \hline
        	2.0 & 40 & 6.6 $\pm$ 0.7 \\
        	2.5 & 45 & 9.0 $\pm$ 1.0 \\
        	3.0 & 50 & 11.0 $\pm$ 1.0 \\ \hline
    	\end{tabular}
    	\caption{HPK HQR75 from~\cite{19-T01007_HPK_SiPMsCharacterizationCT}}
	\end{subtable}
\caption{Experimentally measured cross--talk probabilities at CT.}
\label{tab:x-talk_probs}
\end{table}

The cross--talk correction factors~($\rm f_{XT}$) used for the analysis are computed as follows:
\begin{ceqn}
\begin{equation}
	\rm f_{XT} = \frac{1}{1 + P_{XT}} \pm \frac{\Delta P_{XT}}{\left(1 + P_{XT}\right)^2} \quad .
	\label{eq:x-talk}
\end{equation}
\end{ceqn}

To asses the XA PDE and compare the performances of the different configurations, we choose the bias OV value of 4.5~OV for FBK TT and 3.0~OV for HPK HQR75, since for these values the two models exhibit similar gain (see Figure~\ref{fig:gains_xa}) while keeping the $\rm P_{XT} < 20\%$.

\begin{figure}[htp]    
	\centering
	\begin{subfigure}[b]{0.5\textwidth}
     	\centering
	\includegraphics[height=0.70\linewidth]{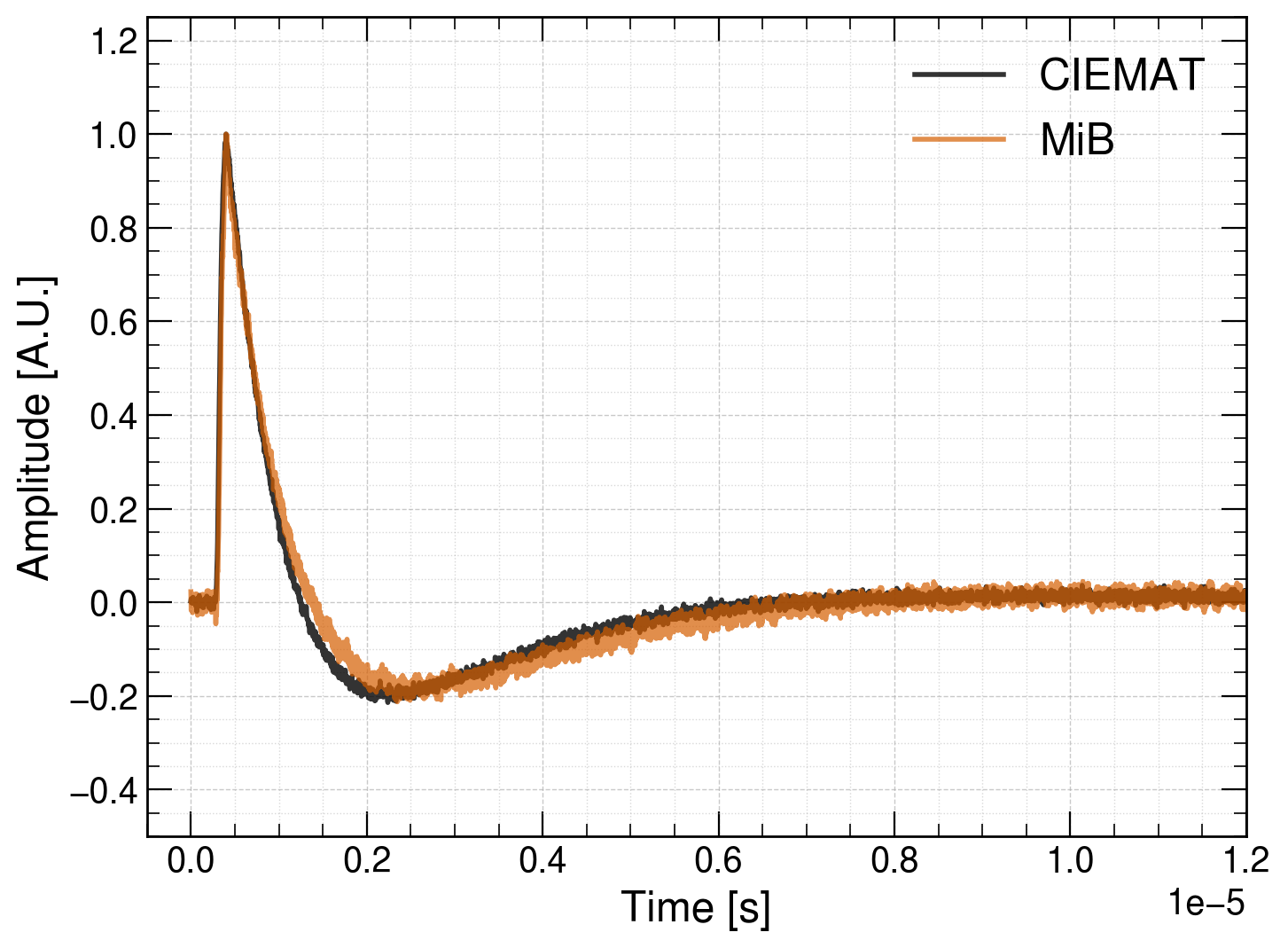}
	\caption{}
	\label{fig:spe_fbk}
	\end{subfigure}%
    
	\begin{subfigure}[b]{0.5\textwidth}
    	\centering
	\includegraphics[height=0.70\linewidth]{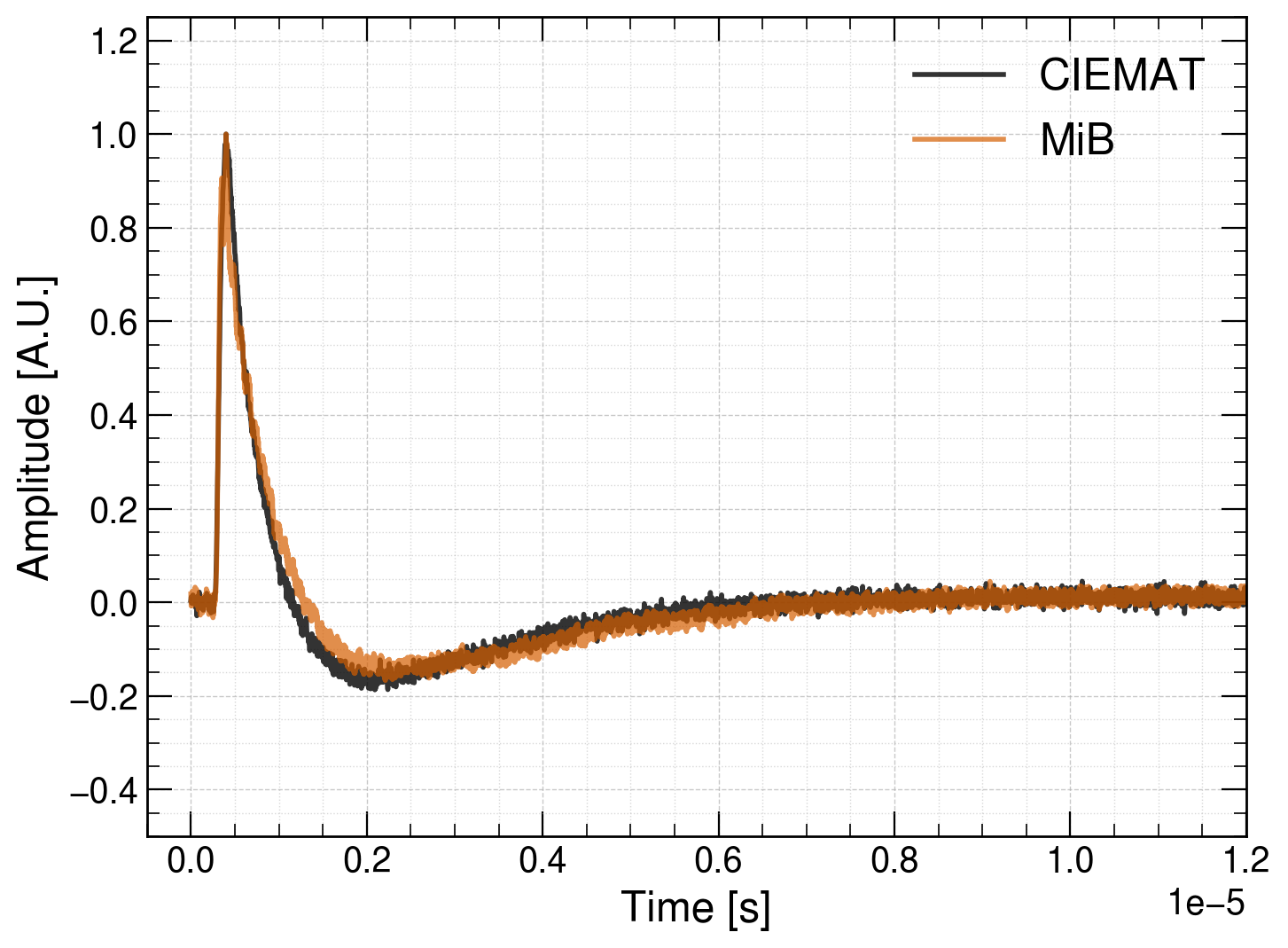}
	\caption{}
	\label{fig:spe_hpk}
	\end{subfigure}
	\caption{Single photo--electron normalized response obtained during calibration in CIEMAT and Milano--Bicocca setups. XAs with (a) FBK and (b) HPK SiPMs.}
\label{fig:spes}
\end{figure}

The average waveforms for the SPE are computed for each XA configuration and the results are shown in Figure~\ref{fig:spes} for both MIB and CIEMAT setups.

The front-end electronics provides  bipolar signals with a characteristic zero crossing time of 1.0~$\mu$s. Due to this behaviour we integrate only the positive part of the signal. The LAr light emission time profile has a triplet component  with characteristic emission time $\tau_t\sim1.6\ \mu s > 1.0\ \mu s$, so to assess the integral of the charge lost in the negative lobe two methods are considered: either the alpha/muon waveforms are deconvolved (CIEMAT) by the SPE response, or the scintillation time profile is convoluted with the SPE response (MiB). The fraction of positive lobe charge is corrected after this computation.

\subsection{Milano Bicocca analysis} \label{sec:ana_mib}

In the method adopted at MiB and used as secondary method by CIEMAT, the XA efficiency ($\epsilon_{\rm MiB} (\rm XA)$) is computed from the ratio of the detected~(\#PE(XA)) to the expected~(\#Ph) light: 
\begin{ceqn}
\begin{equation}
	\epsilon_{\rm MiB} (\rm XA)  = \frac{\#PE(XA)}{\#Ph}\cdot f_{corr} \quad .
	\label{eq:mthb2}
\end{equation}
\end{ceqn}
where \#Ph is:
\begin{equation}
	\#{\rm Ph}  = \rm LY_{LAr}\ E_{\alpha}\ \Omega = 35700\ {\rm ph}/{\rm MeV}\cdot 5.48\ {\rm MeV}\cdot\Omega ,
	\label{eq:mthb1}
\end{equation}
The maximal LAr light yield for alpha particles in LAr $\rm LY_{LAr}=(35700 \pm 2157)$~photons/MeV including the $\alpha$~quenching factor $q_{\alpha} = (0.70 \pm 0.04)$ is from ~\cite{Doke1981,Hitachi1983},  while the geometrical acceptance~($\rm \Omega$) is determined by the Monte Carlo simulations discussed in section~\ref{sec:MC}.  
\#PE(XA) is derived from the fit of the full energy peak of the calibrated alpha spectra as described later in this section.

\begin{figure}[htp]    
	\centering
	\begin{subfigure}[b]{0.5\textwidth}
     	\centering
	\includegraphics[width=\linewidth]{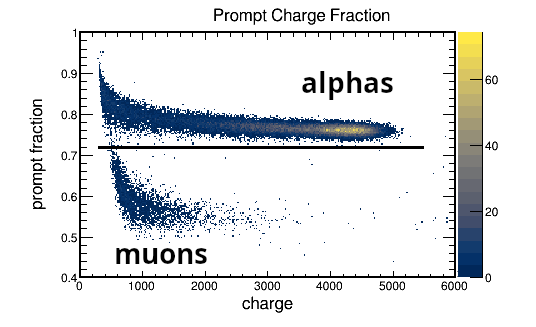}
	\caption{}
	\label{fig:fprompt}
	\end{subfigure}%

	\begin{subfigure}[b]{0.5\textwidth}
    	\centering
	\includegraphics[width=\linewidth]{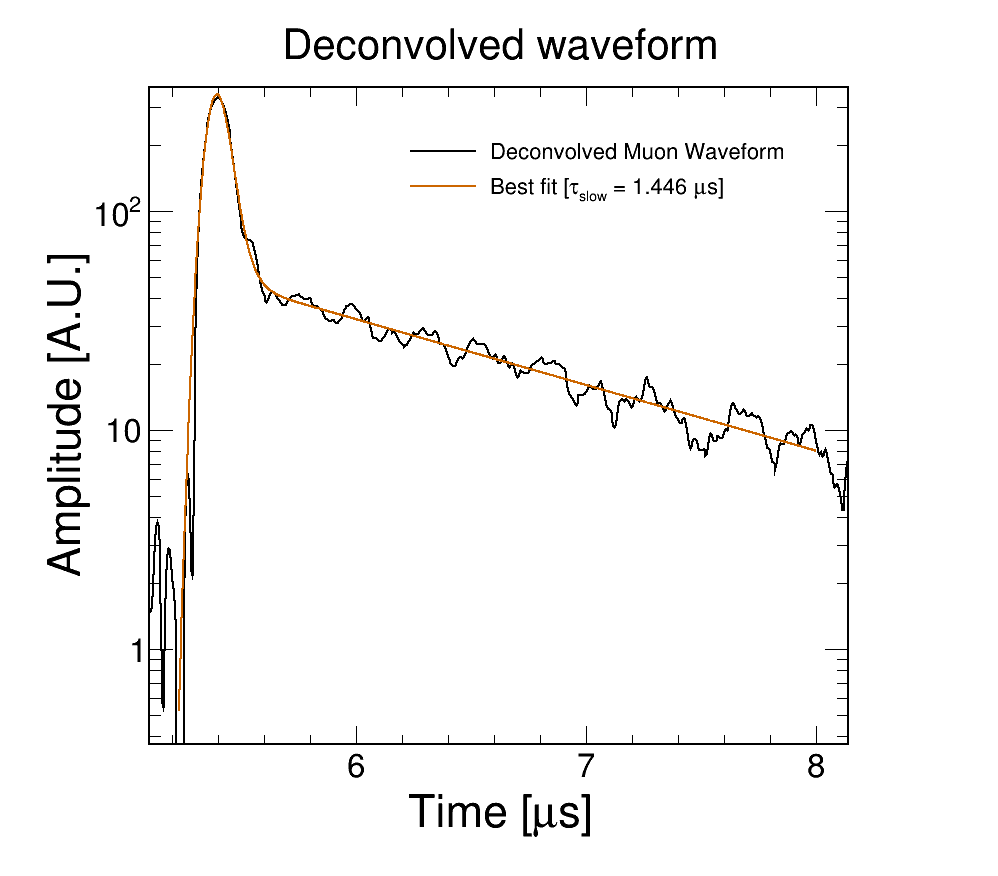}
	\caption{}
	\label{fig:deconvolved_muons}
	\end{subfigure}
	\caption{(a) Fraction of prompt over total integrated charge for an alpha run; alpha and muons populations are clearly distinguishable by cutting on the prompt fraction. (b) The deconvolved muon waveform measured with the MiB setup.}
	\label{fig:muon_selection}
\end{figure}

The alpha spectra are the histograms of the alpha waveforms selected by pulse shape discrimination criteria ~(PSD)~\cite{PDE_XA_JINST} and charge integrated over 1000 ns.
Figure~\ref{fig:fprompt} shows the capability of the MiB setup and method for particle identification by PSD when cutting on the fraction of the prompt~(charge integral $\rm < 600~ns$) over the total (charge integral $\rm < 1000~ns$), named hereafter $\rm f_{prompt}$.  Muons and alphas are clearly separated in the ($\rm f_{prompt}$) vs total charge plane: alphas have $\rm f_{prompt} > 0.7$  and muons $\rm f_{prompt} < 0.7$. More details on the alpha spectra analysis are provided later on in this section.  

The correction factor~($\rm f_{corr}$) is the product
\begin{ceqn}
\begin{equation}
	\rm f_{corr} =  \rm f_{XT} \cdot \rm f_{int} \cdot \rm f_{purity} \quad .
	\label{eq:mthb3}
\end{equation}
\end{ceqn}
and takes into account the cross--talk ($\rm f_{XT}$), the fraction of light falling in the waveform positive lobe ($\rm f_{int}$) and the effective LAr light yield ($\rm f_{\rm purity}$), the latter being related~\cite{Acciarri_2009_N2+O2inLAr} to the light quenching impurity (e.g. N$_2$)  concentration that may vary at each filling of the experimental chamber. 

The correction for the effective LAr yield ($\rm LY_{eff}$) is relevant for the absolute PDE determination and to fairly compare the PDE of the different XA configurations. Up to several ppm values the impurities affect only the triplet (or slow) component of the LAr emission that for alpha particles accounts for only 23\% of the $\rm LY_{LAr}$. The re--normalization factor ($\rm f_{\rm purity}$) accounts for the fraction of the actual triplet ($\rm \tau_{exp}$) to the maximal ($\rm \tau_{pure}$) component~\cite{Acciarri_2009_N2+O2inLAr}
\begin{ceqn}
\begin{equation}
	\rm f^{-1}_{\rm purity} = A_{\rm slow} \frac{\tau_{\rm exp}}{\tau_{\rm pure}} + A_{\rm fast} \quad ,
	\label{eq:purity}
\end{equation}
\end{ceqn}
where $\rm  A_{fast} = 0.77$ and $\rm  A_{slow} = 0.23$ are the literature singlet and triplet contributions for alphas and $\rm \tau_{pure}~=~1600~ns$ is the triplet time constant for the maximal LAr LY~\cite{PDE_XA_JINST,Hitachi1983}.

At each filling of the experimental chamber i.e. for each of the tested XA configurations, the $\rm \tau_{exp}$ is extracted from the muon waveforms analysis. A muon run is taken with the source located at the top position to limit the number of alpha events triggering the DAQ and muons waveforms are selected by PSD criteria: the selected muon waveforms are individually deconvoluted by the SPE waveform template and the resulting normalized muon average waveform is fitted by a two exponential function convoluted with a Gaussian, providing $\rm \tau_{exp}$ as shown in Figure~\ref{fig:deconvolved_muons}. 
The $\rm \tau_{exp}$, hence the quality of the LAr, is monitored and found to be stable along the alpha data taking time.  

The $\rm \tau_{exp}$ measured ranges  are reported in Table~\ref{tab:MiB_LAr_purity} for each XA configuration together with the the corresponding $\rm {f_{int} \cdot f_{purity}}$.
 
\begin{table}[htp]
	\centering
	\begin{tabular}{clcc}
	\hline
	\multicolumn{2}{c}{Measurement} & $\rm \tau_{exp}\ [ns] $ & $\rm {f_{int} \cdot f_{purity}}$ \\ \hline \hline
	(A) & FBK $+$ Eljen	& 910 -- 1113 &  0.843\\
	(B) & FBK $+$ G2P  	& 910 -- 1115 &  0.843\\
	(D) & HPK $+$ G2P  	& 1407 -- 1507 & 0.853\\ \hline
	\end{tabular}
	\caption{The LAr triplet decay time constant $\rm \tau_{exp}$ ranges derived from the muons analysis and the corresponding combined correction factor for the three measurements with the MiB setup, computed for the best fit in the range.}
	\label{tab:MiB_LAr_purity}
\end{table}

\begin{figure}[htp]
    \centering
	   \includegraphics[width=\linewidth]{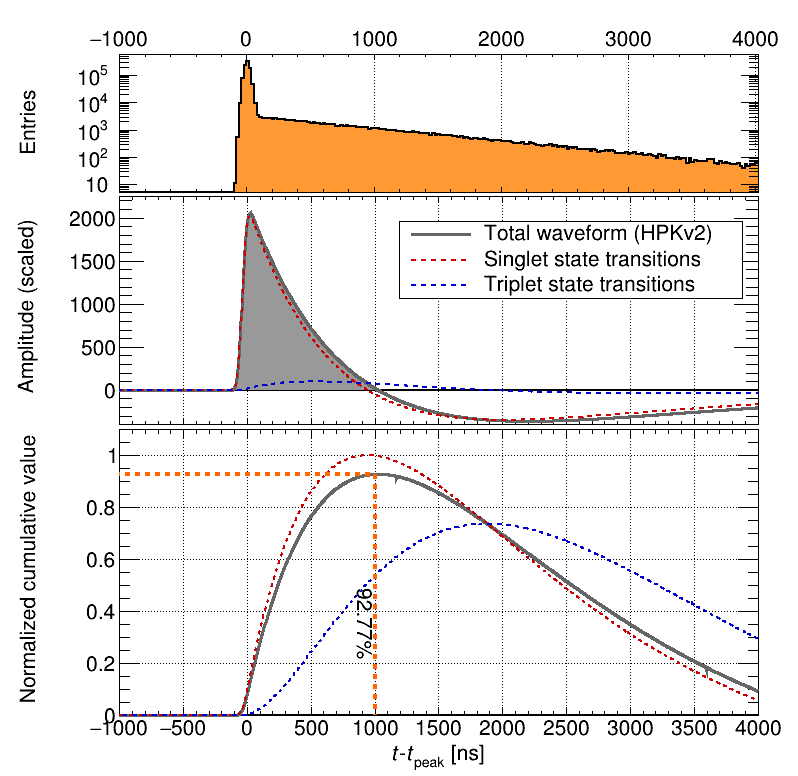}
	   \caption{Computation of the charge integration correction factor for $\tau_{exp} = 963$ ns. From top to bottom: simulated time profile, time profile convolved with electronics response, fraction of integrated charge (y axis) as a function of the integration time (x axis). $\rm {f_{int}}= 92.77\%$  is found  for a $1000$ ns integration window.}
	   \label{fig:f_int}
\end{figure}

The waveform bipolar shape requires to assess the actual charge integrated within the positive lobe of the alpha waveforms ($\rm f_{int}$), that is in turn anti-correlated to $\rm f_{purity}$ and $\rm \tau_{exp}$; in fact at the increase of the latter and since it is greater than the waveform zero crossing time, a larger fraction of the late photons are lost since they fall into the waveform negative lobe. Therefore the $\rm {f_{int} \cdot f_{purity}}$ is numerically computed as follows and described in Figure~\ref{fig:f_int}: 
the alpha particle scintillation time profile for a given $\rm \tau_{exp}$, (top panel), is convoluted with the SPE template and provides the  expected pulse shown in the middle panel. The integration correction factor ($\rm f_{int}$) is finally reported in the Figure~\ref{fig:f_int} (bottom) as a function of the integration time: at each time, it is the ratio of the grey area of the middle panel (the cumulative of the positive lobe of the convoluted waveform) to the orange area of the top panel (the integral of the entire raw scintillation signal). As an example, a $\rm \tau_{exp} = 963$~ns gives $\rm {f_{int}}= 92.77\%$ for a $1000$~ns integration time and $\rm f_{purity} = 90.84\%$. The $\rm f_{int} \cdot f_{purity}$ factor is then applied to the alpha pulse height spectra to properly asses the absolute PDE. The error on $\rm f_{int} \cdot f_{purity}$ is computed by varying $\rm \tau_{pure}$ and $\rm \tau_{exp}$ and found to be $<2\%$; as reported in Table~\ref{tab:MiB_LAr_purity} their product is stable for  $\rm \tau_{exp}$ ranging from 900 to 1500 ns.

\begin{figure}[htp]    
	\centering
	\begin{subfigure}[b]{0.55\textwidth}
    	\centering
	\includegraphics[width=\linewidth]{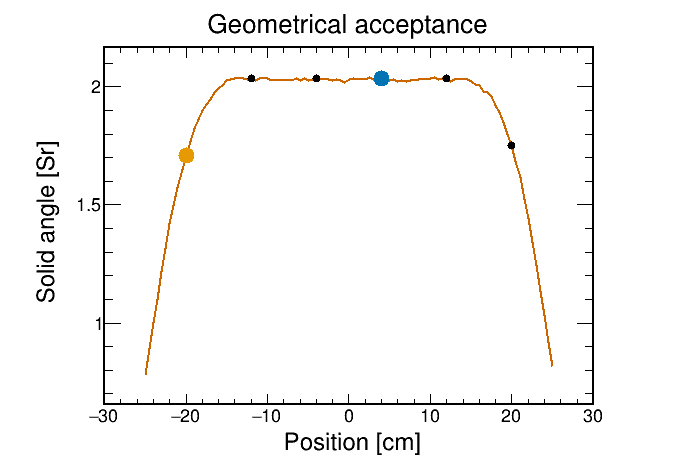}
	\caption{}
	\label{fig:geo_acceptance}
	\end{subfigure}
	\begin{subfigure}[b]{0.55\textwidth}
     	\centering
	\includegraphics[width=\linewidth]{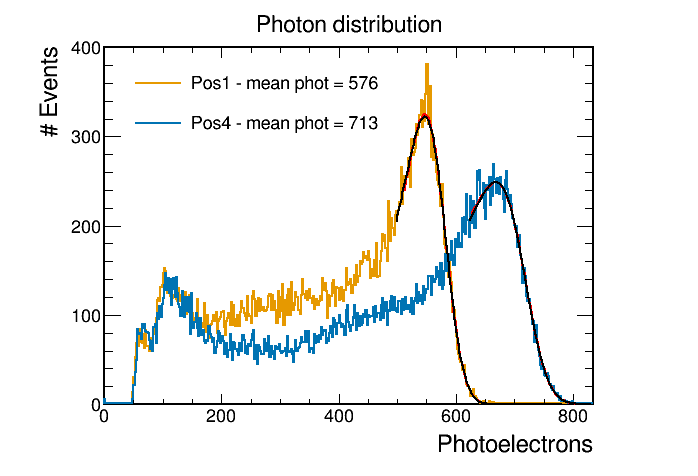}
	\caption{}
	\label{fig:ph_distribution}
	\end{subfigure}%

	\caption{(a) Geometrical acceptance vs source position relative to the XA center. The orange and blue dots correspond to the positions of the source for the 2 spectra shown inpanel b. (b) The alpha spectra measured at the two source positions highlighted in panel a.}
	\label{fig:alphas}
\end{figure}

Figure~\ref{fig:geo_acceptance} shows the module geometrical acceptance as a function of the source position; the values are retrieved by the Montecarlo GEANT4 simulations described in section~\ref{sec:MC}. Two calibrated (by SPE charge) alpha spectra are also shown in Figure~\ref{fig:ph_distribution}; the difference of the detected photoelectron peak values is due to the yet uncorrected geometrical acceptances, that are marked with the same color in the plot above.The tail extending to the low energies comes from degraded light events. The alpha full energy peak is fitted with a gaussian convoluted with an exponential, and the number of detected photons is retrieved by the gaussian mean value.

In the MiB setup, the $^{241}$Am source can slide along a rail mounted in front of the module under test,  allowing to measure the PDE at several positions as shown in the Figure~\ref{fig:MiB_setup_scan}, to study the module uniformity. Figures \ref{fig:F_E_MiB}, \ref{fig:F_G_MiB} and \ref{fig:H_G_MiB} report the measured PDE against the source position for each of the three studied XA configurations. For the positions below $\rm 20$ cm  the PDE looks systematically, despite not significantly, lower than at higher positions. This behaviour is more pronounced for the FBK measurements. All the measurements have been performed with the same module frame and dichroic filters, the only variables being the SiPMs boards (FBK or HPK) and the light guide (G2P or Eljen). For both HPK and FBK, the observed decrease in efficiency at the lowest positions could be explained by a non optimal SiPM-light guide coupling due to the tolerances of both the frame and light guides. The more evident decrease for FBK could be explained by a lower photon detection efficiency of one (few) SiPMs facing that region; this can happen if they have a higher breakdown voltage, hence a lower gain at the same bias voltage. The results obtained with the MiB setup are summarized in Table~\ref{tab:MiB_results}. 

\begin{table}[htp]
\centering
\resizebox{\columnwidth}{!}{%
	\begin{tabular}{ccccc}
	\hline
	(A) & (B) & (C) & (D)   \\ \hline \hline
	1.80 $\pm$ 0.15 & 2.22 $\pm$ 0.19 & - & 2.40 $\pm$ 0.20 \\ \hline 
	\end{tabular}%
}
	\caption{Results for the absolute efficiency ($\rm \epsilon_ {MiB}$(\%)).}
	\label{tab:MiB_results}
\end{table}

The repeated measurements of the HPK \& G2P show a high reproducibility: in Figure~\ref{fig:H_G_MiB}, three independent measurements performed in different days and with two different LAr fillings report that the efficiencies measured at the same positions are well within one standard deviation. This suggests that the main error source are the systematics, as the geometrical acceptance and the cross talk probability uncertainties, while the run to run variance is low. The errors reported in Table~\ref{tab:MiB_results} are to be considered for the absolute PDE, while the relative differences among the tested configurations have lower errors. 

\begin{figure}[htp]    
	\centering
 	\begin{subfigure}[b]{0.40\textwidth}
     	\centering\includegraphics[height=0.33\linewidth]{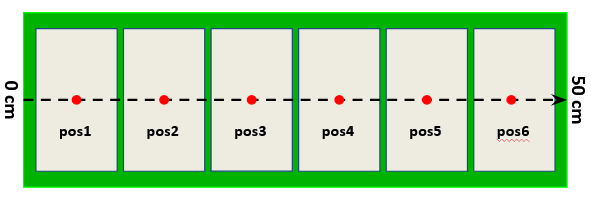}
	\caption{}
	\label{fig:MiB_setup_scan}
 	\end{subfigure}%
\end{figure}%

\begin{figure}[htp]\ContinuedFloat
	\begin{subfigure}[b]{0.50\textwidth}
     	\centering
	\includegraphics[height=0.68\linewidth]{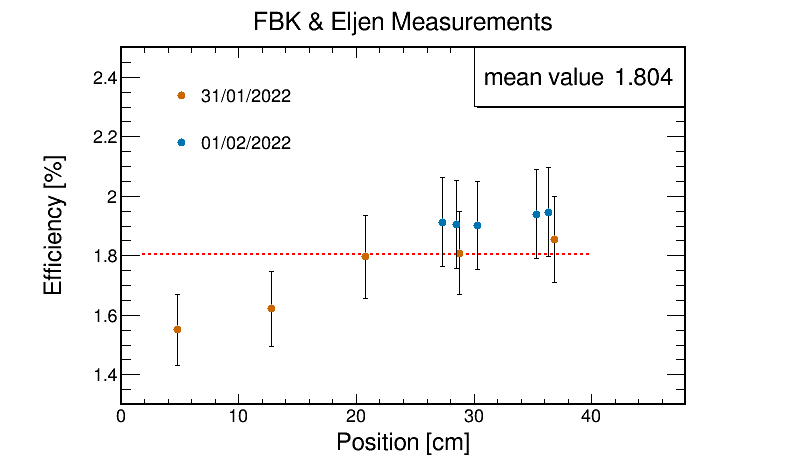}
	\caption{}
	\label{fig:F_E_MiB}
	\end{subfigure}%
    
	\begin{subfigure}[b]{0.50\textwidth}
    	\centering
	\includegraphics[height=0.68\linewidth]{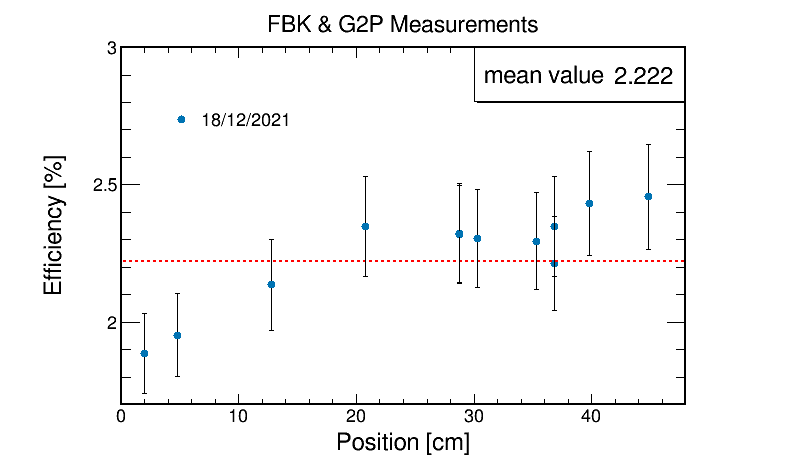}
	\caption{}
	\label{fig:F_G_MiB}
	\end{subfigure}
    
	\begin{subfigure}[b]{0.50\textwidth}
    	\centering
	\includegraphics[height=0.68\linewidth]{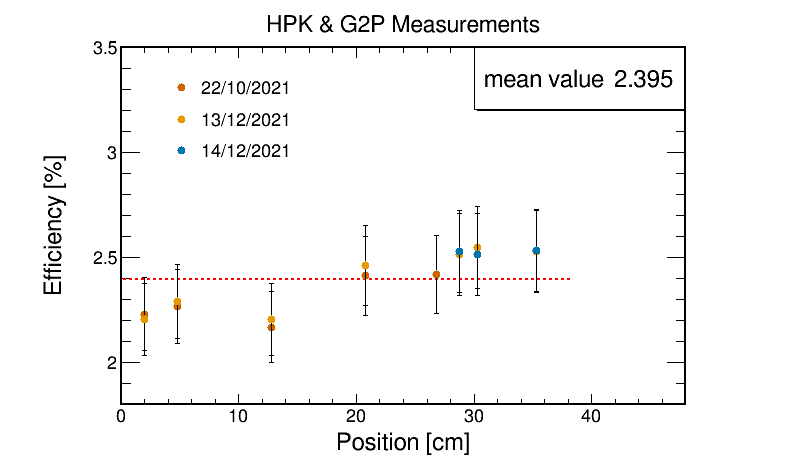}
	\caption{}
	\label{fig:H_G_MiB}
	\end{subfigure}%
    
	\caption{(a) Scheme of the XA irradiation positions. (b, c, d) The measured PDE for vs the source positions, for different device configurations (HPK--FBK and G2P--Eljen). }
	\label{fig:mib_results}
\end{figure}

\newpage
\subsection{CIEMAT analysis} \label{sec:ana_ciemat}

In the method used by CIEMAT, the efficiency of the XA, $\epsilon_{\rm MAD} (\rm XA)$, is obtained from the known efficiency of the reference SiPM’s, $\epsilon (\rm Ref. SiPM)$, as is shown in
equation~\ref{eq:mtha}.
\begin{multline}
	\epsilon_{\rm MAD} (\rm XA) = \frac{\#PE_{u. a.}(XA)}{\#PE_{u. a.}(Ref. SiPM)}\cdot \epsilon (Ref. SiPM) \cdot \\ \cdot \frac{\rm \Omega_{u. a.}(XA)}{\rm \Omega_{u. a.}(Ref. SiPM)}\cdot \rm f'_{corr}   
	\label{eq:mtha}
\end{multline}
\noindent where, u.a is used for the variables computed per unit area, $\rm \#PE_{u.a}$ is the number of PE and $\rm f'_{corr}$ is a correction factor that includes the cross--talk correction ($\rm f_{XT}$) and the fraction of light not included in the waveform integration ($\rm f'_{int}$). The different responses to the extension of the source depending on the sensitive area of the detector are included in $\left[\rm \Omega_{\rm u. a.}(\rm XA)/\rm \Omega_{\rm u. a.}(\rm Ref. SiPM)\right]$. This geometrical acceptance ($\rm \Omega$) is obtained with the Monte Carlo introduced in section~\ref{sec:MC}.

\begin{figure}[htp]
	\centering
	\includegraphics[width=\linewidth]{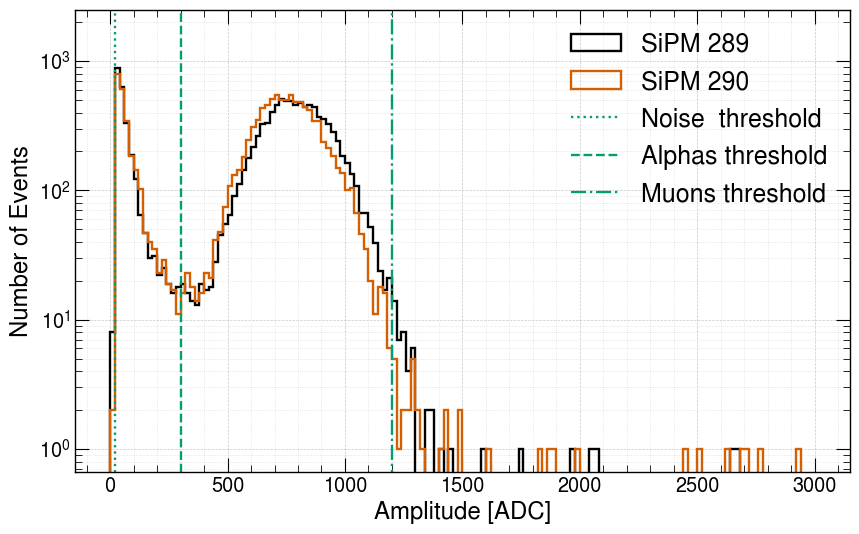}
	\caption{Amplitude histogram for both reference VUV4 SiPM where we can see the different thresholds used for acquiring noise (22~ADC), alpha (300~ADC) and muons (12000~ADC) events. }
	\label{fig:sipms_tr}
\end{figure}

The acquisition in the CIEMAT setup is triggered when the amplitude of both reference SiPMs is above a certain threshold. In Figure~\ref{fig:sipms_tr} the amplitudes of the two SiPMs are shown for a threshold of 1~PE (22 ADC); the light from the $\alpha-$source is clearly distinguishable and allows to set the threshold for triggering $\alpha$ events as marked in the plot (300 ADC). The trigger rate is about 26~Hz in agreement with half of the source activity (considering 2$\pi$ $\alpha$ emission). Due to the small size of the box and the alpha source rate, the number of muon events is small enough 
not to interfere with the alpha signal although they are filtered by amplitude.

\begin{figure}[htp]
	\centering
	\includegraphics[width=\linewidth]{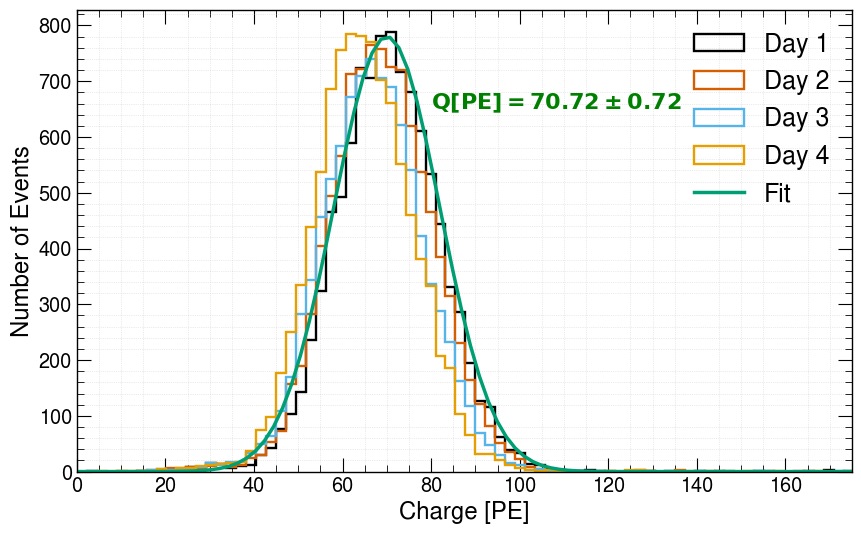}
	\caption{FBK+EJ XA $\alpha-$spectrum for OV 4.5~V and the different days of data taking. Additionally, a gaussian fit for the first day is shown. }
	\label{fig:xa_Q}
\end{figure}

The $\alpha$--spectrum detected by the XA upon the SiPMs triggering is displayed in Figure~\ref{fig:xa_Q}. These distributions are fitted to Gaussian functions, estimating the number of detected PEs as the mean of the distribution.

\begin{figure}[htp]
	\centering
	\includegraphics[width=\linewidth]{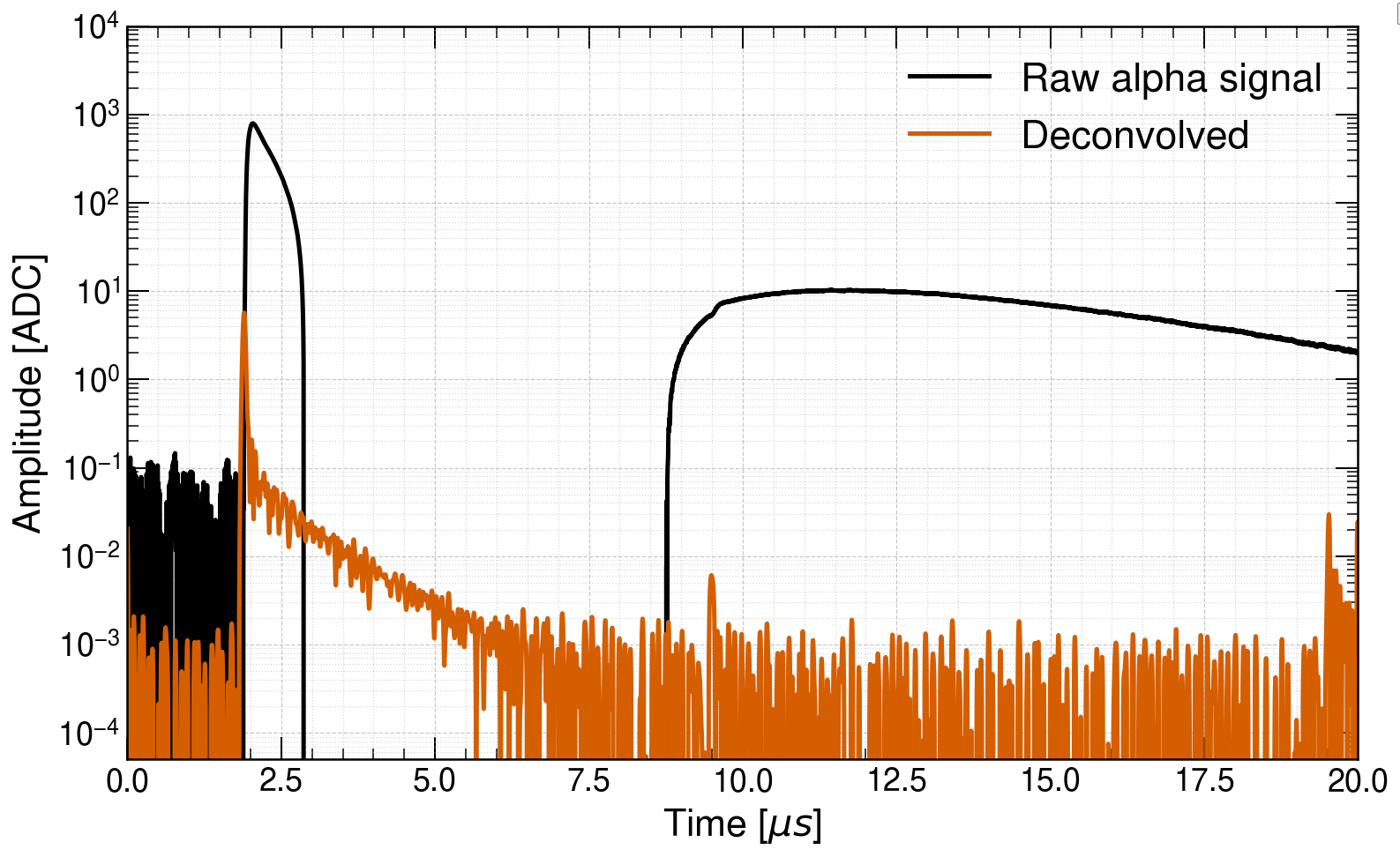}
	\caption{XA raw and deconvolved average waveform of alpha signal for the HPK+G2P configuration. The deconvolution process remove the undershoot that is part of the XA response leaving the expected LAr scintillation profile.}
	\label{fig:HPK+G2P_deconvoluted}
\end{figure}

In this case, the XA plus electronics response is  deconvolved of the light signal profile to determine the full charge and compute $\rm f'_{int}$. The pure XA response is obtained from a 405~nm laser signal; a Gaussian filter is applied in the Fourier space to remove high frequency components. An example of a deconvolved waveform is shown in Figure~\ref{fig:HPK+G2P_deconvoluted}.

By comparing the integrated charge before and after the deconvolution is performed we obtain:
\begin{ceqn}
\begin{equation}
	\rm f'_{int} = 1.09 \pm 0.02 \quad .
	\label{eq:integration}
\end{equation}
\end{ceqn}

This factor depends on the purity of the setup and on the transfer function of the cold--electronics readout. For CIEMAT setup, the achieved purity for the different measurements remains stable enough to assume an unique correction factor for all the configurations. The deviations are included in the error quoted in equation~\ref{eq:integration}.
The final correction factor also includes the effect of the cross--talk probability as defined in equation~\ref{eq:x-talk}:
\begin{ceqn}
\begin{equation}
	\rm f'_{corr} \equiv \frac{f_{XT} (\text{XA})}{f_{XT} (\text{Ref.SiPM})} \ f'_{int} \quad .
\end{equation}
\end{ceqn}
CIEMAT setup allows to compute the absolute efficiency of the XA by a direct comparison between the detected number of PE per mm$^2$ of the XA and the HPK VUV4 reference SiPMs.
As presented in equation~\ref{eq:mtha}, the exposure dimensions are different depending on the photosensor and so are the corresponding solid angles. The results are summarized in Table~\ref{tab:solid_angle}.

\begin{table}[htp]
\centering
\resizebox{\columnwidth}{!}{%
	\begin{tabular}{ccc}
	\hline
 	& XA & Reference SiPM  \\ \hline \hline
	$\rm \Omega$~(Sr) &  0.29 $\pm$ 0.02 & 0.034 $\pm$ 0.003  \\
	Effective area (mm$^2$) &  415.47 & 36.00 \\
	$\rm \Omega_{\rm u.a.}$ &  6.9 $\pm$ 0.5 & 9.4 $\pm$ 0.8 \\ \hline
	\end{tabular}%
}
	\caption{Results for the geometrical acceptance of the XA and the VUV4 reference SiPMs in [Sr] and per unit area (u.a.).}
	\label{tab:solid_angle}
\end{table}

The ratio of the geometrical acceptances per unit area is:
\begin{ceqn}
\begin{equation}
	\frac{\rm \Omega_{\rm u.a.}(\text{Ref.SiPM)}}{\rm \Omega_{\rm u.a.}\text{(XA)}} = 1.35 \pm 0.15 \quad .
	\label{eq:geometrical}
\end{equation}
\end{ceqn}

Following equation~\ref{eq:mtha} we have obtained values for the efficiency of the XA for each day and campaign with respect to each SiPMs.

The final values presented in  Table~\ref{tab:ciemat_A_results} are computed with a weighted average of the values of each day and for each reference SiPM.

\begin{table}[htp]
\centering
\resizebox{\columnwidth}{!}{%
	\begin{tabular}{cccc}
	\hline
	(A) & (B) & (C) & (D)   \\ \hline \hline
	1.34 $\pm$ 0.24 & - & 1.59 $\pm$ 0.29 & 2.13 $\pm$ 0.38  \\ \hline
    	\end{tabular}%
}
	\caption{Results for the absolute efficiency ($\rm \epsilon_ {MAD}$(\%)) with CIEMAT method.}
	\label{tab:ciemat_A_results}
\end{table}

The main source of systematics is the reference VUV4 SiPM PDE ($14\%$), followed by the geometrical acceptance ($10.8\%$).

\begin{figure}[htp]
	\centering
	\includegraphics[width=\linewidth]{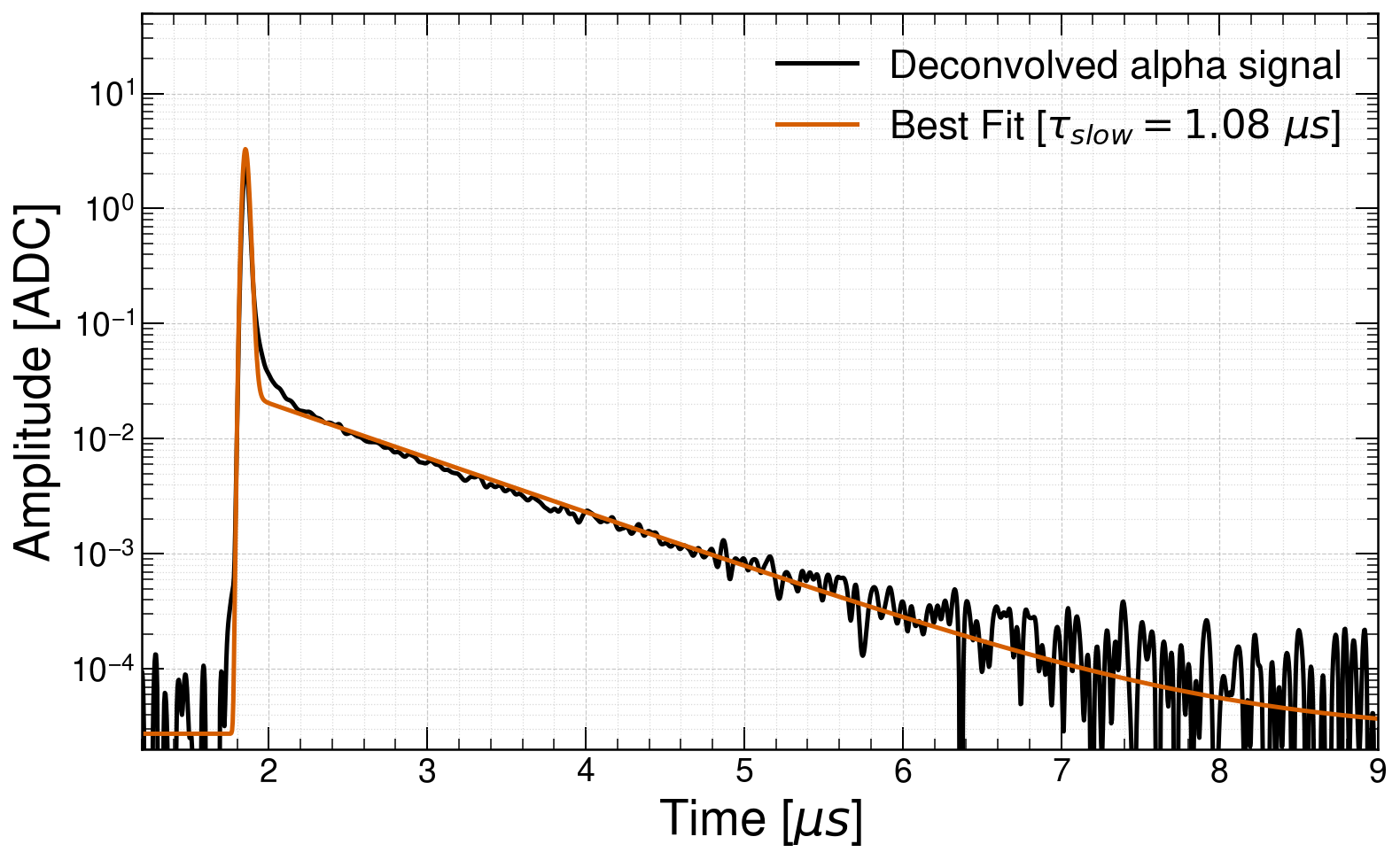}
	\caption{Alpha VUV4 SiPM scintillation profile fitted with a two--exponential function used to extract the $\rm \tau_{slow}$ parameter.}
	\label{fig:sipm_scintillation}
\end{figure}

The efficiency has also been computed from the $\alpha$--source light--yield
that it is used as a cross--check (see equation~\ref{eq:mthb2}). 
As explained in section~\ref{sec:ana_mib}, an additional correction accounting for the quenching of the impurities is needed. CIEMAT uses the scintillation profile provided by the reference SiPMs for $\alpha-$triggered events, as it is shown in Figure~\ref{fig:sipm_scintillation}. The purity in our setups is good enough to assume linear dependence between the charge and the value of $\tau_{\text{slow}}$. In the CIEMAT setup the variation of the purity beyond the linear range can be determined through its effect on the collected charge by the reference SiPMs.

The obtained results following this approach are summarized in Table~\ref{tab:ciemat_B_results}.

\begin{table}[htp]
\centering
\resizebox{\columnwidth}{!}{%
	\begin{tabular}{ccccc}
	\hline
	(A) & (B) & (C) & (D)   \\ \hline \hline
	1.61 $\pm$ 0.12 & -- & 1.86 $\pm$ 0.15 & 2.50 $\pm$ 0.21  \\ \hline
	\end{tabular}%
}
	\caption{Results for the absolute efficiency ($\rm \epsilon'_ {MAD}$(\%)) with the MiB method.}
	\label{tab:ciemat_B_results}
\end{table}

\section{Results} \label{sec:results}

In this section we summarize the absolute efficiency results for the XA  with the two different methods applied. In the first~($\rm \epsilon_{MIB}$), the number of detected photons are compared to the Monte Carlo estimates for the described geometry, a PDE scanning along the cell is perfomed and the average value is provided; for the second one, ($\rm \epsilon_{MAD}$) CIEMAT group made a direct comparison with the reference VUV4 SiPMs known efficiency and measure the PDE at one spot. Additionally, CIEMAT used MiB method as crosscheck ($\rm \epsilon'_{MAD}$) For this purpose, we take the results from sections~\ref{sec:ana_mib} and~\ref{sec:ana_ciemat} (Tables~\ref{tab:MiB_results},~\ref{tab:ciemat_A_results} and~\ref{tab:ciemat_B_results}) where we have calibrated and computed the number of detected photo-electrons respectively.

\begin{table}[H]
\centering
\resizebox{\columnwidth}{!}{%
    \begin{tabular}{ccccc} 
    \hline
    & FBK + EJ & FBK + G2P & HPK + EJ & HPK + G2P \\ \hline \hline
    $\rm \epsilon_{MIB}$  & 1.80 $\pm$ 0.15 & 2.22 $\pm$ 0.19 & - & 2.40 $\pm$ 0.20 \\
    $\rm \epsilon_{MAD}$  & 1.34 $\pm$ 0.24 & - & 1.59 $\pm$ 0.29 & 2.13 $\pm$ 0.38  \\ 
    $\rm \epsilon'_{MAD}$  & 1.61 $\pm$ 0.12 & - & 1.86 $\pm$ 0.15 & 2.50 $\pm$ 0.21  \\
     \hline 
    \end{tabular}%
}
    \caption{Results for the XA absolute efficiency ($\epsilon_i$(\%)).}
    \label{tab:results}
\end{table}

\section{Conclusions} \label{sec:conclusions}
In this article, we report the  measurement of the absolute detection efficiency of the XA in the dedicated cryogenic setups of MiB and CIEMAT. The two setups attain high purity and reproducible results. Two methods with completely different systematic uncertainties ensure the reliability of the measurement.
The XA is the basic unit of the PDS of the first module of the DUNE FD (FD1) and its basic performance and collection efficiency needs to be addressed before its installation in ProtoDUNE, where its global performance will be studied for the first time.
Four different XA configurations exist, in terms of SiPM model and WLS bar manufacturer; two of these element combinations were measured in the two labs, providing compatible results within 1$\sigma$.
The PDE has been measured to be about 2\% for a SiPM bias voltage corresponding to a gain of about $2.5\ 10^{6}$ electrons for both models, i.e., an OV of 3~V for HPK and 4.5~V for FBK SiPMs.
The lower value of the PDE is found for the combination FBK and EJ (1.34 -- 1.80)\%. The PDE is 10\% higher for HPK SiPMs and the same WLS bar. 
This is under investigation and maybe related to the 
smaller active area of the FBK sensor (32.8~mm$^2$ versus the 36~mm$^2$ of HPK).   
The PDE increases between 24\% and 35\% with the G2P WLS bar.
These measurements show a clear better performance of the XA equipped with G2P bar and call for an optimization of the SiPM and WLS bar optical coupling.
Forty units of each of the four tested configurations are installed in ProtoDUNE--HD in the NP04 facility at the CERN neutrino platform, and they will be operated in 2024.

\section*{Acknowledgments} \label{sec:Acknowledgments}

The present research has been supported and funded by INFN in the framework of the DUNE project, the European Union NextGenerationEU/PRTR, European Union’s Horizon 2020 Research and Innovation programme under Grant Agreement No. 101004761, by MCIN/AEI/10.13039/501100011033 under Grants No. PID2019--104676GB--C31, RYC2021--031667--I, PRE2019--090468 and PRE2020--094863 of Spain. Work produced with the support of a 2023 Leonardo Grant for Researchers in Physics, BBVA Foundation.

\clearpage
\bibliography{references.bib} 

\end{document}